\documentclass[12pt]{oxarticle}
\pdfoutput=1

\usepackage{graphicx, float, array, xspace, amscd, amsmath, amsthm, amssymb, latexsym, relsize, bbm}
\usepackage[a4paper, left=1.6in, right=0.3in, top=0.7in, bottom=1.1in]{geometry}
\usepackage[bbgreekl]{mathbbol}
\usepackage{amsfonts, tikz-cd}
\usepackage[T1]{fontenc}

\DeclareSymbolFontAlphabet{\mathbb}{AMSb}
\DeclareSymbolFontAlphabet{\mathbbl}{bbold}

\usepackage[sort&compress, comma, square, numbers]{natbib}
\usepackage[all,cmtip]{xy}
\usepackage{color}
\definecolor{MyDarkBlue}{rgb}{0.15,0.25,0.45}

\usepackage[linktocpage=true,hypertexnames=false]{hyperref}
\hypersetup{colorlinks=true, citecolor=blue, linkcolor=blue, urlcolor=blue, pdfauthor={}, pdftitle={},breaklinks=true}

\let\SS=\S 

\newcommand{\LC}{\text{\tiny LC}}
\newcommand{\Hu}{{\text{\tiny H}}}
\newcommand{\Bi}{{\text{\tiny B}}}

\def\Dbar{{\overline{D}}}

\renewcommand{\sb}{{\overline{\sigma}}}

\newcommand{\kb}{{\overline{ \kappa}}}
\newcommand{\rb}{{\overline{ \rho}}}

\newcommand{\Kb}{{\overline{ K}}}

\newcommand{\w}{{\,\wedge\,}}
\newcommand{\wt}{\widetilde}
\newcommand{\wh}{\widehat}

\newcommand{\fD}{{\mathfrak{D}}}

\newcommand{\half}{\frac{1}{2}}

\def\CS{{\text{CS}}}

\newcommand{\ab}{{\overline\alpha}}
\newcommand{\bb}{{\overline\beta}}

\renewcommand{\a}{{\alpha}}

\renewcommand{\d}{\delta}\newcommand{\D}{\Delta}

\renewcommand{\th}{\theta}\newcommand{\Th}{\Theta}

\renewcommand{\k}{\kappa}
\renewcommand{\l}{\lambda}
\newcommand{\m}{\mu}
\newcommand{\n}{\nu}

\renewcommand{\r}{\rho}
\newcommand{\s}{\sigma}\renewcommand{\S}{\Signa}
\renewcommand{\t}{\tau}

\renewcommand{\o}{\omega}\renewcommand{\O}{\Omega}

\DeclareFontFamily{OT1}{pzc}{}
\DeclareFontShape{OT1}{pzc}{m}{it}{<-> s * [1.200] pzcmi7t}{}
\DeclareMathAlphabet{\mathpzc}{OT1}{pzc}{m}{it}

\newcommand{\cA}{\mathcal{A}}
\newcommand{\ccB}{\mathpzc B}

\newcommand{\cD}{\mathcal{D}}\newcommand{\ccD}{\mathpzc D}

\newcommand{\cF}{\mathcal{F}}

\newcommand{\cH}{\mathcal{H}}

\newcommand{\cM}{\mathcal{M}}

\newcommand{\cO}{\mathcal{O}}

\newcommand{\cR}{\mathcal{R}}

\newcommand{\ccZ}{\mathpzc Z}

\newcommand{\ccZb}{{\overline \ccZ}}

\DeclareFontFamily{U}{bbold}{}
\DeclareFontShape{U}{bbold}{m}{n}
 {  <-5.5> s*[1.05] bbold5
    <5.5-6.5> s*[1.05] bbold6
    <6.5-7.5> s*[1.05] bbold7
    <7.5-8.5> s*[1.05] bbold8
    <8.5-9.5> s*[1.05] bbold9
    <9.5-11.5> s*[1.05] bbold10
    <11.5-16> s*[1.05] bbold12
    <16-> s*[1.05] bbold17
 }{}

\newcommand{\IR}{\mathbbl{R}}

\newcommand{\IZ}{\mathbbl{Z}}

\font\eightrm=cmr8 at 8pt
\font\csc=cmcsc10

\newcommand{\beq}{\begin{equation}}
\newcommand{\eeq}{\end{equation}}
\newcommand{\beqnn}{\begin{equation*}}
\newcommand{\eeqnn}{\end{equation*}}
\newcommand{\bea}{\begin{eqnarray}}
\newcommand{\eea}{\end{eqnarray}}
\newcommand{\bean}{\begin{eqnarray*}}
\newcommand{\eean}{\end{eqnarray*}}

\newcommand{\sref}[1]{\SS\ref{#1}}

\newcommand{\ee}{\text{e}}
\newcommand{\ii}{\text{i}}

\newcommand{\place}[3]{\vbox to0pt{\kern-\parskip\kern-7pt
                             \kern-#2truein\hbox{\kern#1truein #3}
                             \vss}\nointerlineskip}

\DeclareFontFamily{U}{wncy}{}
\DeclareFontShape{U}{wncy}{m}{n}{<->wncyr10}{}
\DeclareSymbolFont{mcy}{U}{wncy}{m}{n}
\DeclareMathSymbol{\sha}{\mathord}{mcy}{"58}

\newcommand{\del}{{\partial}}
\newcommand{\delb}{{\overline{\partial}}}

\newcommand{\lb}{{\overline\lambda}}
\newcommand{\nb}{{\overline\nu}}
\newcommand{\mb}{{\overline\mu}}

\newcommand{\A}{\cA}
\renewcommand{\aa}{\mathfrak{a}}

\newcommand{\dd}{{\text{d}}}

\newcommand{\K}{K\"ahler\xspace}

\newcommand{\tr}{\text{Tr}\hskip2pt}

\newcommand{\tb}{{\overline{\tau}}}

\newcommand{\ap}{{\a^{\backprime}\,}}
\renewcommand{\rb}{{\overline{\rho}}}
\renewcommand{\=}{\;=\;}

\makeatletter
\g@addto@macro\bfseries{\boldmath}
\makeatother

\newcommand{\citeE}{\cite{McOrist:2021dnd}\xspace}

\newcommand{\citeM}{\cite{Candelas:2016usb}\xspace}

\newcommand{\citeSG}{\cite{McOrist:2019mxh}\xspace}
\newcommand{\citeMatter}{\cite{McOrist:2016cfl}\xspace}
\newcommand{\citeUG}{\cite{Candelas:2018lib, McOrist:2019mxh}\xspace}

\renewcommand{\baselinestretch}{1.1}
\numberwithin{equation}{section}
\proofmodefalse
\begin{document}
\pagestyle{empty}      

\begin{center}
\null\vskip0.2in
{\Huge A Heterotic Hermitian--Yang--Mills Equivalence \\[0.5in]}

{\csc Jock McOrist$^{*\,1}$, Sebastien Picard$^{\sharp\,2}$ and
Eirik Eik Svanes$^{\dagger \,3}$\\[0.2in]}

{\it $^*$Department of Mathematics\hphantom{$^2$}\\
School of Science and Technology\\
University of New England\\
Armidale, 2351, Australia\\[3ex]

$^\sharp$ Department of Mathematics\\
University of British Columbia\\
 1984 Mathematics Road\\
  Vancouver, BC, Canada\\[3ex]
  
$^\dagger$ Department of Mathematics and Physics \\
Faculty of Science and Technology\\
University of Stavanger\\
N-4036, Stavanger, Norway\\
}

\footnotetext[1]{{\tt  jmcorist@une.edu.au} 
\hspace*{3.21cm}$^2\,${\tt spicard@math.ubc.ca} \hfil
\hspace*{3.21cm}$^3\,${\tt eirik.e.svanes@uis.no}}
%
\vspace{0.3cm}

\vfill
{\bf Abstract\\[-8pt]}
\end{center}

We consider $N=1$, $d=4$ vacua of heterotic theories in the large radius limit in which $\ap \ll 1$. We construct a real differential operator $\mathcal{D}= D+\bar{D}$ on an extension bundle $(Q, \cD)$ with underlying topology $Q=(T^{1,0}X)^* \oplus {\rm End} \, E \oplus T^{1,0} X$ whose curvature is holomorphic and Hermitian-Yang-Mills with respect to the complex structure and metric on the underlying non-\K complex 3-fold $X$ if and only if the heterotic supersymmetry equations and Bianchi identity are satisfied. This is suggestive of an analogue of the Donaldson--Uhlenbeck--Yau correspondence for heterotic vacua of this type.

\vskip100pt
\newgeometry{left=1.5in, right=0.5in, top=0.75in, bottom=0.8in}
%
\newpage
{\baselineskip=10pt\tableofcontents}
\restoregeometry
%
%
\newpage
\setcounter{page}{1}
\pagestyle{plain}
\renewcommand{\baselinestretch}{1.3}
\null\vskip-10pt


\section{Introduction}

We want to understand the parameter space of a heterotic string compactification to four-dimensional Minkowski spacetime with minimal supersymmetry at large radius. Large radius means a limit in moduli space where all curvatures measured in some appropriate length scale are small, so that heterotic supergravity is a good approximation to the string theory. In this case, supergravity is really an effective field theory for string theory with $\ap$ measuring how close the approximation is to string theory. We want $\ap\ll1$ in order to ensure our supergravity theory is not wallowing in the swampland. In \cite{Hull:1986kz,Strominger:1986uh}, the requirements of realising $N=1$ and $d=4$ Minkowski space were phrased in terms of differential geometric equations on the geometric data which we  summarise shortly and have collectively become known as the Hull-Strominger equations.\footnote{These papers worked to first order in $\ap$ only. How these equations are modified to higher orders in $\ap$, equivalently derivatives, is an open question. Partial progress has however been made, see e.g. \cite{Bergshoeff:1989de}.} 
 The advantage of this formulation is that supersymmetry is directly related to the differential geometry of the underlying geometry of the string compactification. The geometry is non--\K and yet has a \K parameter space, which by itself is mathematically surprising. Indeed, there is now significant mathematical interest in the geometry and parameter space of these compactifications. 
 
 In previous work \citeE we described how to reformulate the conditions for the vacuum to be supersymmetric in terms of the kernel of a nilpotent operator $\Dbar$ and its adjoint $\Dbar^\dag$ . The former corresponded to the system satisfying a set of `F-term' equations as they also can be derived from a superpotential functional while the latter we called `D-term' equations. Indeed, the D-terms  look closely related to those  derived from a moment map construction as is expected from general supersymmetry lore \cite{Ashmore:2019rkx, Garcia-Fernandez:2020awc}. The terminology thence is chosen to mirror the corresponding constructions in $N=1$ $d=4$ supersymmetric field theories and the $\Dbar$--operator seems to play a role analogous to a BRST operator. 
 
The $\Dbar$ operator is interesting as it is not a connection on a vector bundle, but instead the $\ap$--corrections give it a non--trivial derivative structure. It is, nonetheless, a nilpotent operator on a bundle and our work here suggests it should be thought of as a generalisation of a holomorphic structure. 

This paper starts with inspiration from the Hodge dual relation between $\delb^\dag$ and $\del$ in $\delb$--cohomology: we construct a $D$--operator based on the adjoint $\Dbar^\dag$ computed in \citeE. It is nilpotent and allows us to form a real operator $\cD= \Dbar + D$. This again is not a connection, but it does have a  curvature $\cF$ which we compute. What we demonstrate is that this curvature is holomorphic and satisfies a Hermitian-Yang-Mills equation if and only if the supersymmetry and Bianchi identity are satisfied. This sets us on the path to reformulating the existence of heterotic vacua as not so much satisfying the Strominger-Hull equations, but finding holomorphic HYM operators on a certain vector bundle. In other words, our works suggests there might be a generalisation of the Donaldson--Uhlenbeck--Yau theorem for the existence of heterotic vacua at large radius. 

The idea of reinterpreting solutions to the Strominger--Hull equations as a sort of instanton connection on an extension bundle $Q$ was introduced in  \cite{delaOssa:2017pqy,delaossa:2017restrictions}, and this point of view was further developed to define a notion of stability associated to the system by  \cite{garcia2023,garcia202b}. Our approach is different, as we remove the spurious modes related to a Hermitian-Yang-Mills connection $\theta$ on $T^{1,0}X$ which is present in the aforementioned works. In doing this, one particular difference that arises is that $\bar{D}$ in \cite{garcia2023,garcia202b} defines a holomorphic structure, while the $\bar{D}$-operator appearing here is a differential operator which does not define a connection due to additional $\ap$-corrections.

\subsection{Hull--Strominger as a description of heterotic at large radius}
We now summarise Hull-Strominger equations for supersymmetry.  The spacetime geometry is taken to be of the form 
$$
\IR^{3,1} \times X~.
$$
The manifold is complex, which means it admits an integrable complex structure $J = J_m{}^n \dd x^m \otimes \del_n$, where $x^m$ are real coordinates on $X$.  The complex structure is integrable meaning $J^2 = -1$ with its Nijenhuis tensor vanishing. We often have need for wedges -- in the interest of clutter reduction, we  omit the wedge symbol, writing for example $\dd x^{mn} \cong \dd x^m \w \dd x^n$. Occasionally, there may be ambiguity in which we case  we write the wedge explicitly.  

The complex manifold has trivial canonical bundle and so admits a holomorphic volume form $\Omega$. It admits a Riemannian metric $\dd s^2 = g_{mn} \dd x^m \otimes \dd x^n$, which is compatible with the complex structure. It follows that there exists a hermitian 2-form $\o = \half \o_{mn} \dd x^{mn} = \ii g_{\m\nb} \dd x^{\m\nb}$, where we have evaluated the hermitian form in the complex coordinates that diagonalise a given $J$. 

There is a vector bundle $E \to X$, with a connection $A$, and a structure group being a subgroup of $E_8\times E_8$ or ${\rm Spin}(32)/\IZ_2$. The gauge connection is anti-hermitian meaning
$$
A \= \A - \A^\dagger~,
$$
where $\A$ is the $(0,1)$ component of $A$. The field strength for the connection is  $F= \dd A + A^2$ and supersymmetry requires it be holomorphic $F^{(0,2)} = 0$ and  solve a Hermitian Yang--Mills 
$$
g^{mn} F_{mn} \=0~,
$$
which on $X$ can be alternatively written as $\o^2 F = 0$.     

There is a 3-form $H$ which is defined on the one hand as the contorsion of $\o$
\beq\label{eq:ddco}
H \= \dd^c \o \= \frac{1}{3!} J^m J^n J^p (\dd \o)_{mnp}~.
\eeq	
For fixed complex structure this is $H=\ii (\del-\delb)\o$. 
The hermitian form is conformally balanced
$$
\dd \left(e^{-2 \Phi} \o^2\right) \=0~,
$$	
where $\Phi = -{1 / 2} \log |\Omega|_\omega$. We remark that in the case of $H=\cO(\ap)$, which covers the vacua considered in this document, the dilaton $e^{-2 \Phi}$ turns out to be constant to this order in $\ap$ \cite{Anguelova:2010ed}, so the metric is balanced in the sense of Michelsohn \cite{michelsohn82}, but we will not make direct use of this fact as the dilaton will not play a prominent role in our analysis. 

On the other hand the Green--Scharwz anomaly together with coupling of supergravity to a gauge field forces $H$ to be  
\beq
H \= \dd B - \frac{\ap}{4}\Big(\CS[A] - \CS[\Th]\Big)~,
\label{eq:Hdef}\eeq
where $\CS$ is the Chern--Simons form, $B$ is the B-field, and $\Th^\Hu$ is the Hull connection: $\Th^\Hu = \Th^\LC - \half H$.  $H$ satisfies a Bianchi identity
\beq\label{eq:Anomaly0}
\dd H ~=- \frac{\ap}{4} \Big( \tr (F^2) - \tr (R^{\Hu\,2})\, \Big)~.
\eeq
Note there is no ambiguity in the Bianchi identity in evaluating the curvature for $R^\Hu$ if we want our equations to approximate the string theory moduli space. There are other systems of equations one can study in which $R$ in this equation is evaluated with respect to another connection, for example Chern, but this is a different mathematical problem, not relevant to string theory at large radius in which $H=\cO(\ap)$. String theory at large radius already has lots of interesting mathematics to be understood such as the generalisation of mirror symmetry, and so is our focus.

The role of $\ap$--corrections, or quantum corrections, are important physically and mathematically. The Hull--Strominger equations describe the requirements of $N=1$ $d=4$ supersymmetry correct to first order in $\ap$. With the Bianchi identity, these imply the equations of motion, summarised in \sref{app:sugra}, are satisfied, again to first order in $\ap$ \cite{amp24, LopesCardoso:2003dvb, Martelli:2010jx}. The converse is not true, nor is it obvious what the analogous equations are to higher orders in $\ap$. At $\ap^2$ for example, it is likely the manifold is no longer conformally balanced and so the role of $\ap$--corrections is important as far the differential geometry of these string theory vacua go. Furthermore, non--perturbative effects in $\ap$ play an important role. These are computed via the string world sheet as instantons and normally carry information about geometric invariants.  For example, in the study of Calabi--Yau manifolds and associated $(2,2)$ CFTs it was only after including the $\ap$--corrections did one discover mirror symmetry: that the moduli space of a CY manifold $X$ and its mirror $\hat X$ are the same after a change of coordinates. The $\ap^3$ correction to the \K potential carries information about the Riemann--Zeta function.  This term can be understood as a consequence of mirror symmetry. The corrections also encode geometric invariants such as Gromov--Witten invariants, see e.g. \cite{Candelas:1990qd,Candelas:1990rm} and \cite{Morrison:1994fr}.  For heterotic theories discussed here, we know the equations of motion up to $\ap^2$, but the  $\ap^3$ are not yet fixed \cite{Anguelova:2010ed,Melnikov:2014ywa}. That being so, it is important (if we want to relate our work to string theory, which we do) that we keep track of the order in $\ap$ in which we work, and understand that equations such  as the conformally balanced equation no longer hold at order $\ap^2$. We will at points need to work order-by-order in $\ap$ \cite{Witten:1985bz,Witten:1986kg}. When we do so we will say as much.

\subsection{Moduli of heterotic vacua}
We are interested in deformations of solutions. Deformations of the fields can be related to coordinates of the moduli space via derivatives. In the ten-dimensional physical string theory, this corresponds to a choice of Kaluza-Klein anstaz. 

The gauge field $A$ is antihermitian and so we write it as $A = \A - \A^\dag$ where $\A = A^{(0,1)}$. A deformation is
\beq\label{eq:Avar}
 \A \to \A + \aa ~, \qquad \aa \= \d y^a \fD_a \A + \d y^a \d y^b \fD_a \fD_b \A + \cdots~,
\eeq
In this paper we restrict to first order deformations and so $\aa = \d y^a \fD_a A$.  The derivative is necessary as background gauge transformations on $A$ depend on parameters \citeM. The choice of covariant derivative determines the map between $\d y^a$ and the deformation $\aa$. To save clutter we   write $\aa$ understanding its a small deformation.

A deformation of complex structure can be phrased in many ways. One of the roads to Rome is in terms of a deformation of complex coordinates on $X$:
\beq
\dd x^\m \to \dd x^\m + \d y^a \D_{a\,\nb}{}^\m \,\dd x^\nb~.
\eeq
The index $a$ runs over all possible deformations, for which at the moment, we are not specific about. The $\d y^a$ are tangent vectors to the moduli space $\cM$. Sometimes its convenient to write $\D = \d y^a \D_{a\,\nb}{}^\m \dd x^\nb\otimes \del_\m$. The holomorphic top form $\O$ transforms correspondingly
\beq\label{eq:OmVar}
\O \to \O + \D^\m\, \O_\m~.
\eeq
Requiring the complex structure remain integral means $\delb \D^\m= 0$. 

The B-field transforms under $E_8 \times E_8$ gauge transformations as well as being a gerbe.  There is, nonetheless, a gauge invariant first order deformation $\ccB_a$\footnote{This is most easily derived by differentiating $H$ once: $\d H = \d y^ a\dd \ccB_a + \cdots$} and it naturally combined with a deformation of the hermitian form $\d \o = \d y^a \del_a \o$ to form 
\beq\label{eq:Zvar}
\ccZ_a \= \ccB_a + \ii \del_a \o~, \qquad \ccZb_a \= \ccB_a - \ii \del_a \o~.
\eeq
the generalisation of the complexified \K form for Calabi-Yau manifolds. 

The moduli space is a complex \K manifold and so coordinates split accordingly 
$$
y^a \= (y^\a, y^\bb)~. 
$$
The deformations above admit small gauge transformations, small diffeomorphisms and small gerbes, for example $\aa \to \aa + \delb_\A \phi$. This symmetry is fixed by a choice of gauge; a convenient one is holomorphic gauge \citeSG  in which a subset of the field deformations depend holomorphically on parameters: 
\beq\label{eq:Gauge}
\begin{split}
 \fD_\ab \A &\= 0~, \quad \D_{\ab\nb}{}^\m \= 0~, \\
  \quad  \ccZ_\ab^{(1,1)} &\= 0~, \quad \ccZ_\ab^{(0,2)} \= \ccZb_\ab^{(0,2)} \= \ccZb_\ab^{(2,0)} \= 0~.
\end{split}
\eeq
If we additionally impose that $\d \O^{(3,0)} = k(y) \O$, where $k(y)$ is a parameter dependent constant on $X$ then the gauge, diffeomorphism and gerbe symmetries are fully fixed. 
The physical degrees of freedom are thence
\beq\label{eq:DOF}
\aa ~,\quad \D^\m~,\quad    \ccZ^{(1,1)}~,
\eeq
where from now on we will  suppress the subscript $\a$ and drop the superscript $(1,1)$, except to observe that
$$
 \ccZ_\a^{(1,1)} \= 2\ii (\del_\a \o)^{(1,1)} = 2 \ccB_\a^{(1,1)}~, \qquad \ccZb_\a^{(0,2)} {\=}2 \ccB_\a^{(0,2)} {\=}\! -2\ii (\del_\a \o)^{(0,2)}~.
 $$
 It is of note that $(\del_\a\o)^{(0,2)} = \D_\a{}^\m \o_{\m} = \ii \D_{\a\mb\nb}\dd x^{\mb\nb}$ cannot be gauge fixed to zero. At the standard embedding, so that $H=0$ vanishes, one can show that $(\del_\a\o)^{(0,2)} =0$. For a generic heterotic theory with $H=\cO(\ap)$, it can be shown that $(\del_\a\o)^{(0,2)} = \cO(\ap)$ and so $\D_{\a\mb\nb}$ carries an antisymmetric component. 

We end this section by noting that the study of the moduli of heterotic vacua has been an active field of research which can be understood from various points of view; see e.g. \cite{Anderson:2014xha, delaOssa:2014cia, Garcia-Fernandez:2015hja, Ashmore:2018ybe, Ashmore:2019rkx, Candelas:2016usb, Candelas:2018lib, Garcia_Fernandez_2022, Garcia-Fernandez:2020awc, Garcia-Fernandez:2015hja, Ashmore:2023vji, picardwu} and references therein for recent developments.

\newpage
\section{The extension bundle \texorpdfstring{$Q$}{Q}, the \texorpdfstring{$\Dbar$}{Dbar}--operator and supersymmetry}

A $N=1$, $d=4$  supersymmetric field theory divides up the off-shell fluctuations into F-terms, derivable from a functional the superpotential and D-terms, related to a moment map action on the fields. In string theory, we are always on-shell, but nonetheless we talk about a superpotential for the effective field theory of a compactification to $d=4$ with $N=1$ supersymmetry \cite{Gukov_2000}.  This is a functional of the fields on the $d=6$ compact manifold, and deformations of these fields are required to be holomorphic functions of the parameter space. It is constructed to reproduce a subset of the supersymmetry equations. For heterotic theories are large radius, the functional takes the form \cite{LopesCardoso:2003dvb}:
\beq
\label{eq:W}
W \= \int_X \O\, (H + \ii \dd \o)~.
\eeq
In holomorphic gauge, the functional $W$ is a holomorphic function of parameters \citeMatter in that $\del_\ab W = \del_\ab\del_\bb W = \cdots =  0$ where $y^\a,y^\bb$ are the complex coordinates for the moduli space $\cM$, as is required by $N=1$, $d=4$ supersymmetry. In addition we require $W=0$ and its first holomorphic derivative to vanish  $\del_\a W=0$. It can be checked that the superpotential gives the  Yukawa couplings among the matter and moduli fields that matches a $d=10$ dimensional reduction \citeMatter while  substituting \eqref{eq:Avar}-\eqref{eq:Gauge} into $W$ gives the F-terms \cite{LopesCardoso:2003dvb,Gurrieri_2004,delaOssa:2015maa, Ashmore:2018ybe}:
\beq\label{eq:moduliEqnHol}
\begin{split}
&\delb\, \D^\m \= 0~, \\[0.15cm]
&\delb_\A \aa +  F_\m \w \D^\m\=0~, \\
& \delb \wt\ccZ^{(1,1)}  - 2\ii \, \D{}^\m (\del\o)_\m  -\frac{\ap}{2} \tr \big( F \aa \big) - \ap  (\nabla^\LC_\r \D^\l) R^\r{}_{\l\,\n}  \= 0~,
 \end{split}
\eeq
where we have used the relation derived in \citeUG between the connection $\th$ on $T_X$ with the metric and complex structure data:
\beq\label{eq:dthp}
\d \th^{\n}{}_\m \=  \nabla^\LC_\m  \D{}^\n + \half \, \nabla^{\LC\,\n} \, \ccZ_{\m\rb} \dd x^\rb ~.
\eeq
Substituting into $\tr R\w\d\th$ we get two terms  one of which is absorbed by the redefinition 
\beq
\wt \ccZ_{\m\nb}\= \ccZ_{\m\nb}- \frac{\ap}{2} \nabla^{\LC\,\r} \nabla^\LC_\m  \ccZ_{\r\nb}~,
\eeq
while the second appears on the right hand side of the last line of \eqref{eq:moduliEqnHol}. 
A first order variation of the Hermitian Yang-Mills equation can be written as \footnote{We are free to use $\ccZ$ in lieu of $\wt \ccZ$ as the HYM equation is normalised with a factor of $\ap$. \cite{Witten:1986kg} }
\beq
\label{eq:adjoint2}
\delb_\A^\dag \aa +\half F^{\m\nb} \ccZ_{\m\nb}     \= 0~,
\eeq

A first order variation of the balanced equation is
\begin{align}\notag
\partial(\omega\,\ccZ)&\=0~,\\
-i\delb(\omega\,\ccZ)+2\,\del(\omega{\D}^\m\,\omega_\m)&\=0~.\notag
\label{eq:defbal2}
\end{align}
The first equation is co-closure 
\begin{equation}
\label{eq:coclosedZ}\
\delb^\dagger\ccZ_\a^{(1,1)} \= \delb^\dag \wt\ccZ^{(1,1)}_\a + \frac{\ap}{2}  R^{\t\sb\l\nb} \nabla_\l \wt\ccZ_{\a\,\t\sb} \=0~,
\end{equation}
while the second equation, when combined with the third line of \eqref{eq:moduliEqnHol}, is 
\begin{equation}\label{eq:adjoint1}
\begin{split}
& \frac{1}{2}\,H^{\n\r\mb} \wt\ccZ_{\a\,\r\mb} +\frac{\ap}{4}\Big[\tr\big(F^{\n\mb} \,\fD_\a \A_{\mb}\big)  \Big] - \,\nabla^{{\rm Ch/H}\, \mb}\Delta_{\a\mb}{}^\n - \ii \D_{\a\,\rb\lb} (\del \o)^{\n\rb\lb}  \\
&+  \frac{\ap}{2}   R^\n{}_{\r}{}^{\s\tb} \nabla_\s\D_{\a\,\tb}{}^\r   - \frac{\ap}{4} R^{\n\rb\s\tb} \nabla_\rb \ccZ_\a{}_{\s\tb}{}\=0~,
\end{split}
\end{equation}
We call these equations D-terms as they are in the kernel of an operator $\Dbar^\dag$, where the adjoint is constructed with the moduli space metric derived form a dimensional reduction of the ten-dimensional supergravity action \citeE. Note, the D-terms in this language do not exactly correspond to first order variations of the balanced and HYM equation: we need to use part of the F-term viz., the third line of \eqref{eq:moduliEqnHol}. A similar point was used in  \cite{Ashmore:2019rkx} in the context of a moment map construction in generalised geometry. As an aside, in \citeE this analysis was also undertaken for the non--physical problem in which $\th$ is a connection for $T_X$ not fixed in terms of the metric and other fields. While non-physical, the calculations are arguably cleaner, if not simpler.

\subsection{F-terms and D-terms as kernel of \texorpdfstring{$\Dbar$}{Dbar} and \texorpdfstring{$\Dbar^\dag$}{DbarDag}}
\label{s:Dbar}
In \cite{McOrist:2021dnd} the F-terms and D-terms were phrased as the kernel of  an operator $\Dbar$ and its adjoint $\Dbar^\dag$ respectively. These operators  act on a complex vector bundle
\beq
Q \= (T^{1,0}X)^* \oplus {\rm End} \, E \oplus T^{1,0}X~.
\eeq
Indeed, the idea of encoding the heterotic Bianchi identity into a $\bar{D}$ operator goes back to  \cite{delaOssa:2014cia}.\footnote{See \cite{bismut1989, gualtieri2014} for early instances of this idea in the mathematics literature.} In \cite{McOrist:2021dnd}, this construction was adapted so that ${\rm Tr} \, R^2$ is computed with respect to the metric tensor $g$ rather than spurious field $\Theta$ and the operator there squares to zero up to $O(\alpha'^2)$ corrections, consistent with the fact the Strominger-Hull equations, cf \eqref{susy-1storder}, also suppress $\ap^2$ corrections. We review this work introducing a new feature in that the operator \cite{McOrist:2021dnd} can modified to square to zero exactly provided one assumes a modified Bianchi identity. The distinction between these Bianchi identities in $\ap^2$, so this does not affect the physics but makes for more pleasing mathematics. Indeed, we will show that $\Dbar^2 = 0$ if and only if $(\omega,A)$ satisfies the equation
\beq
i \partial \bar{\partial} \omega \= \frac{\alpha'}{2} {\rm Tr} \, F^2 - \frac{\alpha'}{2} {\rm Tr} \, R^2~.
\eeq
where $R$ is the Chern connection of $\omega$.

First, we establish some notation and definitions that we will use in the coming sections. We often need to use forms valued in vector bundles, for example  $\D \in \Omega^{0,p}(T^{1,0}X)$ is
\beq
\D \= \frac{1}{p!}  \D^\mu{}_{\k_1 \cdots \k_p} \, \partial_\mu \otimes \dd x^{\k_1\cdots\k_p} \= \D^\m{}_K\, \del_\m \otimes \dd x^K ~,
\eeq
or sometimes just $\D = \D^\mu \partial_\mu$. We   use $K$ to as a multi-index, for example $\dd x^K = \dd x^{\k_1} \wedge \cdots \wedge \dd x^\k_k$.

Denote sections $q$ of $\Omega^{(0,p)}(Q)$ by
\beq
q \= \begin{bmatrix} ~\ccZ~ \\ \aa \\ \D \end{bmatrix} \in \begin{bmatrix}~ \Omega^{(0,p)}((T^{1,0}X)^*)~ \\ \Omega^{(0,p)}({\rm End} \, E) \\ \Omega^{(0,p)}(T^{1,0} X) \end{bmatrix}~.
\eeq

Consider $\Dbar: \Omega^{(0,p)}(Q) \rightarrow \Omega^{(0,p+1)}(Q)$  given by
\beq \label{Dbar-defn}
\Dbar \= \begin{bmatrix}
\delb & \ap \cF^* & \cH + \ap \cR \nabla \\
0 & \delb_A & \cF \\
0 & 0 & \delb
\end{bmatrix}~,
\eeq
where
\beq
\cF: \Omega^{(0,p)}(T^{1,0}X) \to \Omega^{(0,p+1)}({\rm End} \, E)~, \qquad \cF^*: \Omega^{(0,p)}({\rm End} \, E) \to \Omega^{(0,p)}((T^{1,0}X)^*)~,
\eeq
is defined by
\beq
\cF (\D) \= F_{\m \nb} \, \dd x^\nb \w \D^\m~, \qquad \cF^* (\aa) \= {\rm Tr} \, F_{\m \nb} \, \dd x^\m \otimes \dd x^\nb \wedge \aa~,
\eeq
while 
\beq
\cH: \Omega^{(0,p)}(T^{1,0}X) \to \Omega^{(0,p)}((T^{1,0}X)^*)~, \qquad \cH (\D) \= H_{\r \nb\m} \,\dd x^\r \otimes \dd x^{\nb} \wedge \D^\m~.
\eeq
Finally, we let 
\beq
(\cR \nabla): \Omega^{(0,p)}(T^{1,0}X) \rightarrow \Omega^{(0,p+1)}((T^{1,0}X)^*)~,
\eeq
be given by 
\beq\label{eq:RNab}
(\cR \nabla) \D \=\! - \frac{1}{p!}  R_{\r \mb}{}^\s{}_\lambda \hat{\nabla}_\s \D^\lambda{}_{\kb_1 \cdots \kb_p} \, \dd x^\r \otimes \dd x^{\mb\kb_1\cdots \kb_p} ~,
\eeq
where $R$ is the curvature of the Chern connection $\nabla$ of $g$ and $\hat{\nabla}$ is the Bismut connection (also known as the Strominger-Bismut connection \cite{bismut1989, Strominger:1986uh}, or the Yano  connection \cite{YanoBook}). The definition of $\hat{\nabla}$ in our notation is given in \eqref{nabla-hat}. We take the convention that $\hat{\nabla}_\s$  only acts on the bundle indices of $\D^\m \in \Omega^{(0,p)}(T^{1,0}X)$ and not the $(0,p)$-form indices. Concretely, acting on for example $\D \in \Omega^{0,1}(T^{1,0}X)$, we have
\beq
\hat{\nabla}_\s \D^\lambda{}_{\kb} \= \partial_\s \D^\lambda{}_{\kb} + \Gamma_\s{}^\lambda{}_\m \D^\m{}_{\kb} + H^\lambda{}_{\s \m} \D^\m{}_{\kb}~,
\eeq
where $\Gamma_\m{}^\n{}_\rho = g^{\sb \n} \partial_\m g_{\rho \sb} $ is the Chern connection. To this order in $\ap$, definition \eqref{eq:RNab} is equivalent to the operator in \citeE, but this difference will be important for future work on $\ap^2$ corrections.

The operator $\Dbar$ defined in \eqref{Dbar-defn} is a differential operator of order 1, but it is not a connection, or a traditional holomorphic structure on $Q$ due to the term $\ap \cR \nabla$. It can however be viewed as an $\ap$-corrected holomorphic structure. In \citeE, the operators $\Dbar$ and $\Dbar^\dag$ are defined as connections but on auxiliary bundles that do not correspond to the physical degrees of freedom in the heterotic theory. The drive to understand physical moduli spaces leads us to consider a more interesting geometric structure. 
\footnote{A curious aside is to consider a non--physical setup in which $\cR \nabla$ in \eqref{eq:RNab} is defined such that $R_{\m \nb}{}^\s{}_\lambda$ and ${\nabla}$ are computed with respect to a  metric $\tilde{g}$, distinct from the metric on $X$. The metric is chosen such that its Chern connection $\tilde{\nabla}$ is a Hermitian-Yang-Mills connection with respect to $(TX,\omega)$. The Bianchi identity \eqref{eq:Anomaly0} is modified by replacing the $\ap^2$-correction $\tr R^\Hu \w R^\Hu$ with a term  ${\rm Tr} \, R_{\tilde{\nabla}} \wedge R_{\tilde{\nabla}}$, in which  $\tilde{\nabla}$ is a Hermitian Yang--Mills connection. This suggests an alternate setup to understanding the equations with a HYM connection for ${\rm Tr} \, R_{\tilde{\nabla}} \wedge R_{\tilde{\nabla}}$; in that case $Q$ is typically \cite{delaOssa:2014cia, Garcia-Fernandez:2015hja, garcia202b} taken as ${\rm End} \, E = {\rm End} \, E_0 \oplus {\rm End} \, T^{1,0}X$, and the current setup removes the spurious summand ${\rm End} \, T^{1,0}X$. \label{ftnote-hym}}

We now compute $\Dbar^2$. For $q \in \Omega^{0,k}(Q)$,  we use $\delb^2 = 0$ and find
\beq
\Dbar^2 q \= \begin{bmatrix} \hat{Z} \\ \delb (\cF \D) + \cF \delb \D \\ 0 \end{bmatrix}~,
\eeq
where
\bea \label{dbar-output}
\hat{Z} &\=& \ap \delb (\cR \nabla \D) + \ap \cR \nabla \delb \D + \ap \delb (\cF^* \aa) + \ap \cF^* \delb \aa \nonumber\\
&&+ \delb (\cH \D) + \cH \delb \D + \ap \cF^* \cF \D~.
\eea
The term $\delb(\cF \D) + \cF \delb \D$ and $\delb (\cF^* \aa) + \cF^* \delb \aa$ in $\hat Z$ both vanish via the Bianchi identity $\dd_A F = 0$ while 
\bea
\delb (\cH \D) + \cH \delb \D + \ap \cF^* \cF \D 
&\=& \bigg( \half (\delb H) + \frac{\ap}{4} ({\rm Tr} \, F \wedge F) \bigg)_{ \r\sb  \m \nb} \dd x^\r \otimes dx^{\sb} \wedge \dd x^{\nb} \wedge \D^\m~. \nonumber
\eea
 Since $\delb H = -i \partial \delb \omega$, we obtain
\beq \label{upsilon-appears}
\Dbar^2 q \= \begin{bmatrix} \half (- i \partial \delb \omega + \frac{\ap}{2} {\rm Tr} \, F \wedge F )_{\r \sb \m \nb} \, \dd x^{\r\sb\nb} \wedge \D^\m - \Upsilon \\ 0 \\0 \end{bmatrix}~,
\eeq
where the main burden of this calculation is in unravelling $\Upsilon$:
\beq 
\Upsilon \= \frac{\ap}{k!} \bigg( \delb_{\sb} ( R_{\r \mb}{}^\t{}_\lambda \hat{\nabla}_\t \D^\lambda{}_{\bar{K}}) - R_{\r \mb}{}^\t{}_\lambda \hat{\nabla}_\t \partial_{\sb} \D^\lambda{}_{\bar{K}} \bigg) \dd x^\r \otimes \dd x^{\sb \mb \bar{K}}~.
\eeq
The Bianchi identity $\delb R=0$ implies  $\partial_{\sb} R_{\r \mb}{}^\t{}_\lambda = \partial_{\mb} R_{\r \sb}{}^\t{}_\lambda$  and so we get
\beq 
\Upsilon \= \frac{\ap}{k!} \bigg(   R_{\r \mb}{}^\t{}_\lambda \partial_{\sb} \hat{\nabla}_\t \D^\lambda{}_{\bar{K}} -  R_{\r \mb}{}^\t{}_\lambda \hat{\nabla}_\t \partial_{\sb} \D^\lambda{}_{\bar{K}} \bigg) dx^\r \otimes dx^{\sb\mb\Kb} ~.
\eeq
Using the definition of $\hat{\nabla}$ in  \eqref{nabla-hat}, we have
\bea
\Upsilon_{\r\sb\mb\Kb} &\=&  R_{\r \mb}{}^\t{}_\lambda \partial_{\sb} (\nabla_\t \D^\lambda{}_{\bar{K}} + H^\lambda{}_{\t \gamma} \D^\gamma{}_{\bar{K}}) - R_{\r \mb}{}^\t{}_\lambda \nabla_\t \partial_{\sb} \D^\lambda{}_{\bar{K}}  -R_{\r \mb}{}^\t{}_\lambda H^\lambda{}_{\t \gamma} \partial_{\sb} \D^\gamma{}_{\bar{K}} \nonumber\\
&\=& R_{\r \mb}{}^\t{}_\lambda [\partial_{\sb},\nabla_\t] \D^\lambda{}_{\bar{K}} + R_{\r \mb}{}^\t{}_\lambda \partial_{\sb} H^\lambda{}_{\t \gamma} \D^\gamma{}_{\bar{K}}~.
\eea
Therefore
\beq
\Upsilon \= \ap  \bigg( - R_{\r \mb}{}^\t{}_\lambda R_{\t \sb}{}^\lambda{}_\gamma +  R_{\r \mb}{}^\t{}_\lambda \partial_{\sb} H^\lambda{}_{\t \gamma}  \bigg) \dd x^\r \otimes \dd x^{\sb\mb}  \wedge \D^\gamma~.
\eeq

 Next, we use the following identity for the curvature of the Chern connection in non-\K geometry:
\beq \label{nonK-curv-sym}
R_{\t \sb}{}^\lambda{}_\gamma \= R_{\gamma \sb}{}^\lambda{}_\t - \partial_{\sb} H^\lambda{}_{\gamma \t}~,
\eeq
which can be seen by direct computation using ({\ref{R-chern-conv}). Therefore
\bea
\Upsilon &\=& \ap \bigg( - R_{\r \mb}{}^\t{}_\lambda R_{\gamma \sb}{}^\lambda{}_\t  \bigg) \dd x^\r \otimes \dd x^{\sb\mb}\w \D^\gamma \nonumber\\
&\=& \frac{\ap}{4} \bigg( - ({\rm Tr} \, R \wedge R)_{\r \mb \gamma \sb} \bigg) \dd x^\r \otimes \dd x^{\sb\mb}\w \D^\gamma~.
\eea
Substituting this into (\ref{upsilon-appears}), we obtain
\beq \label{Dbar-squared-final}
\Dbar^2 q \= \begin{bmatrix} \half (- \ii \partial \delb \omega + \frac{\ap}{2} {\rm Tr} \, F \wedge F - \frac{\ap}{2} {\rm Tr} \, R \wedge R)_{\r \sb \m \nb} \, \dd x^\r \otimes \dd x^{\sb\nb}\w \D^\m \\ 0 \\0 \end{bmatrix}~.
\eeq
This proves the equivalence between $\bar{D}^2=0$ and the truncated 4-form Bianchi identity with Chern connection for ${\rm Tr} \, R \wedge R$. We can substitute the Chern connection curvature $R$ for the Hull connection curvature $R^\Hu$ at the expense of introducing $\ap^2$ corrections since $H = O(\alpha')$. So we see that $\Dbar^2 = 0$ is equivalent to the heterotic Bianchi identity up to $\ap^2$ terms. See also footnote \ref{ftnote-hym} for the construction of a nilpotent operator equivalent to the Bianchi identity with HYM connection on the tangent bundle. 

In \citeE we demonstrate the vanishing $\Dbar$ on $q$ is equivalent to the F-terms derived from the superpotential. Using the moduli space metric constructed for large radius heterotic vacua with $d=4$ $N=1$ supersymmetry in \citeM, \citeUG in which the spurious degrees of freedom are solved for in terms of the metric and complex structure, we construct an adjoint $\Dbar^\dag$ operator. For a section $q \in \Omega^{(0,k)}(Q)$
\beq\label{eq:Dadj}
 \Dbar^\dag  q \= 
 \begin{pmatrix}
\wh \ccZ \\
\wh \aa \\
\wh \D{}^\m
\end{pmatrix}~,
\eeq
where
\beq
\begin{split}
 \wh \ccZ{}^\nb &\=\! -\delb^\dag \ccZ{}^{\nb} + \frac{\ap}{2} (-1)^k R^{\t\sb\l\nb} \nabla_\l \ccZ_{\t\sb}~, \\[3pt]
 \wh \aa& \=  \delb_\A^\dag \aa - \half (-1)^k F^{\r\nb} Z_{\r\nb}~,\\[3pt]
\wh \D{}^\m &\=\! - \frac{(-1)^{k}}{2}H^{\m\r\lb} Z_{\r\lb} + \frac{\ap}{4} \tr (F^{\m\rb}\, \aa_\rb) - \nabla^{{\rm Ch/H}\, \nb}\Delta_{\nb}{}^\m \\
&\qquad\qquad -  \frac{\ap}{2}  (-1)^k R^\m{}_{\r}{}^{\s\tb} \nabla_\s\D_{\tb}{}^\r   + \frac{\ap}{4}(-1)^k R^{\m\rb\s\tb} \nabla_\rb Z_{\s\tb}{}~,\\[3pt]
\end{split}
\eeq	
where in the interest of maintaining semi-compact expressions we have suppressed writing the form indices where possible,  understanding that say  $\aa_\rb \cong \frac{1}{(k-1)!} \aa_{\rb \kb_2\ldots \kb_k} \dd x^{\kb_2\ldots\kb_k}$. The case of interest is $k=1$ and in this case the first and third lines are deformations of the conformally balanced equation, while the middle line is the deformation of the hermitian Yang--Mills equation. That is, $\Dbar^\dag q = 0$ amounts to solving the D-terms: \eqref{eq:adjoint2}, \eqref{eq:coclosedZ}, \eqref{eq:adjoint1} \citeE.

\section{ A \texorpdfstring{$D$}{D}--operator }

Some intuition for the following is to recall that the Dolbeault operator $\del$ on a complex Riemannian manifold is related via the Hodge dual to the adjoint $\delb^\dag$: 
$$
\delb^\dag \=\! - \star \del\, \star~.
$$
Inspired by this, we use  the adjoint $\Dbar^\dagger$   in \sref{s:Dbar}    to motivate a definition for an operator acting in the un-barred directions 
$$
D: \Omega^{k,0}(Q) \rightarrow \Omega^{k+1,0}(Q)~.
$$
 We will then use this to construct a real operator 
$$
\cD \= D +\Dbar~,
$$
and we will demonstrate an equivalence between this operator being Hermitian Yang-Mills and the supersymmetry equations and Bianchi identity. 

\subsection{Motivation from \texorpdfstring{$G_2$}{G2}}

The idea that solutions to the heterotic system on $X$ should solve a heterotic Yang-Mills type condition on an extension bundle $Q$ goes back to  \cite{delaOssa:2017pqy,delaossa:2017restrictions}.  Before delving into the full calculations, we consider the simplified model where we take the heterotic Bianchi identity to be
\beq
dH \= \frac{\ap}{4} {\rm Tr} \, F \wedge F~,
\eeq
and suppress the ${\rm Tr} \, R \wedge R$ term. Let $X$ be a complex 3-fold with complex structure $J$, compatible hermitian metric $g$ and a vector bundle $E \rightarrow X$ together with a connection $A$.  Taking inspiration from the heterotic $G_2$ system discussed in \cite{delaOssa:2017pqy, delaossa:2017restrictions}, we construct the connection
$$
{\cal D} \= 
\begin{pmatrix}
\dd_{\A} & \cF \\
\ap\hat \cF & \nabla^\Hu
\end{pmatrix}~,
$$
on $Q={\rm End}\, E \oplus T_{\mathbb{C}} X$ and let $\nabla^\Hu$ be the Hull connection. 

It is desirable that the on-shell constraints on the geometry can be traced back to the nilpotency of some differential operator, as was the case for heterotic $G_2$ geometries \cite{delaOssa:2017pqy}. Indeed, it is expected that these differentials correspond to BRST operators from the world-sheet point of view.

We would also like to consider the conformally balanced constraint, that is the conformally re-scaled form $\tilde\rho=e^{-2\phi}\omega^2$ is closed, as part of a nilpotency constraint. To do so, consider the following complex
\begin{equation}
    0\rightarrow\Omega^0(Q)\xrightarrow{\tilde\rho\wedge\cal D}\Omega^5(Q)\xrightarrow{\cal D}\Omega^6(Q)\rightarrow0\:,
\end{equation}
where $\tilde\rho=e^{-2\phi}\omega^2$. Requiring this to be a differential complex, that is 
\begin{equation}
    {\cal D}\left(\tilde\rho\wedge{\cal D} f\right)\=0~,
\end{equation}
for all $f\in\Omega^0(Q)$, it is easy to see that this requires both $\dd\tilde\rho=0$ and the Yang-Mills constraint 
\begin{equation}
    \omega^2\wedge{\cal D}^2\=0~,
\end{equation}
on the curvature of $\cal D$.

Given these constraints, that is that $\omega$ is conformally balanced, the differential constraints on $\cal D$ reduce to requiring that $\cal D$ is holomorphic Yang-Mills. We then compute the square of ${\cal D}$ as
$$
{\cal D}^2 \= 
\begin{pmatrix}
\dd_{\A}^2 +\ap \cF\circ \hat \cF  & \dd_{\A,\nabla^\Hu}\cF \\
\ap\dd_{\A,\nabla^\Hu}\hat \cF & \left(\nabla^\Hu\right)^2 + \ap\hat \cF \circ \cF
\end{pmatrix}~.
$$
An order by order argument in $\ap$, similar to the one found in \cite{delaOssa:2017pqy}, shows that imposing the holomorphic and Yang-Mills constraint on the top left corner forces the gauge connection $A$ to be holomorphic and Yang-Mills. This forces the curvature $F$ of the gauge connection to satisfy the instanton condition. The off-diagonal terms readily vanish when imposing the holomorphic Yang-Mills constraint, following a similar argument to the one in \cite{delaOssa:2017pqy}. This leaves the bottom right corner. Using the identity
\begin{equation}
    \hat{R}_{mnpq}-R^{\rm H}_{pqmn} \=\tfrac12(\dd H)_{mnpq}\:,
\end{equation}
the fact that $\dd H$ is of type $(2,2)$ for integrable $J$, and the fact that the curvature of the Bismut connection $\hat{R}$ has frame indices of type $(1,1)$, that is
\begin{equation}
    \hat{R}_{mn\bar a\bar b} \= 0\:,
\end{equation}
imposing the holomorphic constraint forces precicely the Bianchi identity. The Yang-Mills constraint then simply becomes
\begin{equation}
     \hat{R}_{mnpq}\omega^{pq}\=0\:,
\end{equation}
that is, we require the manifold to be Bismut Ricci-flat. In the presence of a holomorphic volume form $\Omega$ on $X$, it is well-known that the Bismut Ricci-flat condition is equivalent to the conformally balanced equation \cite{fino2004}.

We conclude that the above differential constraints on ${\cal D}$ imply that the geometry is conformally balanced, the gauge connection being holomorphic Yang-Mills, and the heterotic Bianchi identity being satisfied. A rewriting of the heterotic action in terms of BPS-spared terms using the almost complex structure as in \cite{LopesCardoso:2003dvb} then shows that the equations of motion are also satisfied.

We remark that this link between a Yang-Mills equation on $Q$ and the Strominger-Hull equations on $X$ was further developed by \cite{garcia2023,garcia202b} where this perspective enabled them to define a Futaki invariant and necessary stability conditions for the Strominger-Hull equations.

The difference between the next section of the current work and  \cite{delaOssa:2017pqy,delaossa:2017restrictions},  \cite{garcia2023,garcia202b}, is that we do not treat the connection on the tangent bundle $\nabla$ as an independent degree of freedom. Rather, the connection $\nabla$ used to compute ${\rm Tr} \, R_\nabla \wedge R_\nabla$ should be determined by the metric tensor $g$. In this work, we use either the Chern connection or the Hull connection $\nabla^H$ to compute ${\rm Tr} \, R_\nabla \wedge R_\nabla$, which are equivalent in the regime $H=O(\ap)$.

We now move on to including the ${\rm Tr} \, R \wedge R$ term in the heterotic Bianchi identity.

\subsection{The \texorpdfstring{$D$}{D} operator for \texorpdfstring{$SU(3)$}{SU(3)} heterotic compactifications}

With \eqref{eq:Dadj} as guiding motivation, the operator $D$ on differential forms valued in the bundle $Q=(T^{1,0}X)^* \oplus {\rm End} \, E \oplus T^{1,0} X$ is defined as
\beq
D  \begin{bmatrix} Z \\ \aa \\ \D \end{bmatrix} \= \begin{bmatrix}
\partial_\nabla + \ap (\cR_1 \nabla) & 0 & 0 \\
 \tilde{\cF} & \partial_\nabla & 0 \\
\tilde{\cH} +  \ap (\cR_2 \nabla) & \ap \tilde{\cF}^* &
\partial_\nabla + \ap (\cR_3 \nabla)
\end{bmatrix} \begin{bmatrix} Z \\ \aa \\ \D \end{bmatrix}~,
\eeq
with
\beq
\tilde{\cF} (Z) \= F_{\r}{}^\m  \, dx^\r \, \wedge Z_\m~, \quad \tilde{\cF}^* (\aa) = {\rm Tr} \, F_\m{}^\r \, \partial_\r \otimes \dd x^\m \wedge \aa~, \quad \tilde{\cH} (Z) = H_\n{}^{\m \r} \, \partial_\r \otimes dx^\n \wedge Z_\m~,
\eeq
and
\bea
(\cR_1 \nabla) Z &\=&\! - \frac{1}{k!} R_\m{}^\s{}^{\lambda}{}_\r \nabla_\lambda Z_{\s K}\, \dd x^\r \otimes \dd x^\m \wedge \dd x^K~, \nonumber\\
(\cR_2 \nabla) Z &\=&\! - \frac{1}{k!} R_\m{}^\r{}^{\s \bar{\lambda}} \nabla_{\bar{\lambda}} Z_{\s K} \, \partial_\r \otimes \dd x^\m \wedge \dd x^K~,\nonumber\\
(\cR_3 \nabla) \D &\=&\! - \frac{1}{k!} R_\m{}^\r{}^\lambda{}_\s \nabla_\lambda \D^\s{}_K \, \partial_\r \otimes \dd x^\m \wedge \dd x^K~.
\eea
Recall our conventions are such that the Chern connection in $\nabla_s Z_{rK}$, $\nabla_s \D^r{}_K$ acts only on the $r$ index, and as the regular derivative on the form index $K$. Similarly,
\beq
\partial_\nabla Z \= \frac{1}{k!} \nabla_\m Z_{rK} \dd x^r \otimes \dd x^\m \wedge \dd x^K~, \qquad \nabla_\m Z_{rK} \= \partial_\m Z_{rK} - \Gamma_\m{}^\lambda{}_r Z_{\lambda K}~,
\eeq
and there is no connection on the form-type $K$ index. We now form the differential operator
\beq
\mathcal{D} \= D + \Dbar~.
\eeq
Let $q \in \Gamma(Q)$, which we write as
\beq
q \= \begin{bmatrix} Z \\ \aa \\ \D \end{bmatrix} \in \begin{bmatrix} \Omega^{1,0} (X) \\ \Gamma({\rm End} \, E) \\ \Gamma(T^{1,0}X) \end{bmatrix}~.
\eeq
Consider
\beq
\mathcal{D}^2 q \= \mathcal{F}_{\mathcal{D}} \, q \= \{ D, \Dbar \} q ~.
\eeq
This should be viewed as an analog of the field strength on $Q$. However, $\mathcal{F}_{\mathcal{D}}$ is not a tensor since it involves derivative operators, though these extra corrections occur at higher order in $\ap$.

To connect with physics, we do as in the introduction and suppose $X$ is a complex manifold of complex dimension 3 with hermitian metric $\omega$ which is Calabi-Yau at zeroth order. This means we can expand fields in a perturbation expansion and consider the equations order-by-order in $\ap$, as done in say \cite{Witten:1986kg}. The metric is 
\beq \label{g-expansion}
g \= g_{\rm CY} + \ap g_1 + O(\ap^2)~,
\eeq
where $g_{\rm CY}$ is a Calabi-Yau metric \cite{yau1978} so that
\beq \label{torsion-alpha}
H \= O(\ap)~.
\eeq

Now consider the following, which we call the heterotic Hermitian Yang-Mills equation,
\beq \label{eq-heterotic-HYM0}
\mathcal{F}_{\mathcal{D}}^{0,2}\=0~, \quad \mathcal{F}_{\mathcal{D}}^{2,0}\=0~, \quad \o^2 \mathcal{F}_{\mathcal{D}}  \= 0~.
\eeq

Our main result is to  establish that for all $Z$, $\aa$, $V$ 
\beq
\Dbar^2 \= 0~, \quad D^2 \begin{bmatrix} Z \\ \aa \\ \D \end{bmatrix} \= \begin{bmatrix} 0 \\ 0 \\ 0 \end{bmatrix} + \begin{bmatrix} O(\ap^2) \\ O(\ap) \\ O(\ap^2) \end{bmatrix}~,
\eeq
and
\beq \label{eq-heterotic-HYM}
g^{\m \nb} [D_\m, D_{\nb}] \begin{bmatrix} Z \\ \aa \\ \D \end{bmatrix} \= \begin{bmatrix} 0 \\ 0 \\ 0 \end{bmatrix} + \begin{bmatrix} O(\ap^2) \\ O(\ap) \\ O(\ap^2) \end{bmatrix}~,
\eeq
is equivalent to
\beq \label{bianchi-1storder}
F^{0,2} \= 0~, \quad i \partial \delb \omega \= \frac{\ap}{2} ({\rm Tr} \, F \wedge F - {\rm Tr} \, R \wedge R)~.
\eeq
and
\beq \label{susy-1storder}
g^{\m \nb} F_{\m \nb} \= O(\ap), \quad \hat{R}_{mn}{}^\m{}_\m \= O(\ap^2)~,
\eeq
where $m,n$ are real indices and $\mu,\nu$ holomorphic indices, and we recall $\hat{R}$ is the curvature of the Bismut connection. If \eqref{bianchi-1storder} is taken with the Chern connection, then the Bianchi identity with Hull connection holds:
\beq \label{bianchi-1storderhull}
i \partial \delb \omega \= \frac{\ap}{2} ({\rm Tr} \, F \wedge F - {\rm Tr} \, R^{\rm H} \wedge R^{\rm H}) + O(\ap^2)~.
\eeq

Solutions to \eqref{torsion-alpha}, \eqref{bianchi-1storder}-\eqref{susy-1storder} can be constructed by perturbing solutions $(\omega_0,A_0)$ at $\ap=0$ where $\omega_0$ is a \K Calabi-Yau metric and $A_0$ is a Yang-Mills connection on a stable bundle $(E,[\omega_0])$ as first described in \cite{Witten:1985bz,Witten:1986kg}. The implicit function theorem then gives solutions $(\omega_{\ap},A_{\ap})$ for $\ap \rightarrow 0$ and the expansion \eqref{g-expansion} is mathematically rigorous \cite{andreas2012,collins2022, liyau2005, Melnikov:2014ywa}.

We note that there is no reason that the physical equations \eqref{susy-1storder}, \eqref{bianchi-1storderhull} should work to all orders in $\ap$. In particular, \eqref{bianchi-1storderhull} should not be exact equality as there is no consistent truncation of heterotic string theory to first order $\ap$ apart from the standard embedding, in which the bundle $E$ is identified with the tangent bundle $T_X$. This is the order to which we work in this paper, demanding that our solutions be viable string solutions, which forces us to large radius, with $H=\cO(\ap)$ so that that there is an underlying sigma model which flows to a $(0,2)$ SCFT \cite{Jardine:2018sft}. Fortunately, this is precisely the regime where the Hull-Strominger equations are valid modulo $\ap^2$ corrections and thence the Hull and Chern connection in the Bianchi identity are indistinguishable as $\tr R\w R$ is normalised with an $\ap$. Of course, extending this to $\ap^2$ is of critical importance, but beyond the scope of this document.

\subsection{The square of \texorpdfstring{$D$}{D}}
Unlike $\Dbar$, the operator $D$ does not square to zero exactly. Instead, in the regime $H = O(\ap)$ it satisfies
\beq \label{DD-identity}
D^2 \begin{bmatrix} Z \\ \aa \\ \D \end{bmatrix} \= \begin{bmatrix} 0 \\ 0 \\ 0 \end{bmatrix} + \begin{bmatrix} O(\ap^2) \\ O(\ap) \\ O(\ap^2) \end{bmatrix}~,
\eeq
if and only if the heterotic Bianchi identity holds. We now verify this. Let $q \in \Omega^{k,0}(Q)$. Then direct calculation gives
\beq \label{DD-direct}
D^2 \begin{bmatrix} Z \\ \aa \\ \D \end{bmatrix} \= \begin{bmatrix} (\partial_\nabla + \ap \cR_1 \nabla)^2 Z \\ \ap \tilde{\cF} (\cR_1 \nabla) Z \\ \hat{\D} \end{bmatrix}~,
\eeq
with
\beq\notag
\begin{split}
 \hat{\D} \=& \tilde{\mathcal{H}} \partial Z + \ap \tilde{\mathcal{H}} (\cR_1 \nabla) Z + \ap \cR_2 \nabla (\partial Z + \ap \cR_1 \nabla Z) \\[2pt]
&+ \ap \tilde{\cF}^* \tilde{\cF} Z + (\partial + \ap \cR_3 \nabla)(\tilde{\mathcal{H}} Z + \ap \cR_2 \nabla Z) \\[2pt]
&+ \ap^2 \cR_3 \nabla \tilde{\cF}^* \aa + (\partial + \ap \cR_3 \nabla)^2 \D~.
\end{split}
\eeq
Here we used $\partial_\nabla^2=0$, $\partial (\tilde{\cF} Z) = - \tilde{\cF}^* \partial Z$. 

We now use the assumption $H = O(\ap)$ and drop terms past leading order in $\ap$. Note that
\beq\notag
(\partial_\nabla + \ap \cR_1 \nabla)^2 Z \= O(\ap^2)~, \qquad (\partial_\nabla + \ap \cR_3 \nabla )^2 \D \= O(\ap^2)~,
\eeq
since
\beq \label{DD-id1}
\partial_\nabla (\cR_1 \nabla Z) + (\cR_1 \nabla) \partial_\nabla Z \= O(\ap)~.
\eeq
To verify \eqref{DD-id1}, we expand it in components which gives
\beq\notag
\bigg( \nabla_\m ( R_\n{}^{\s \lambda}{}_\r \nabla_\lambda Z_{\s K} ) + R_\m{}^{\s \lambda}{}_\r \nabla_\lambda \nabla_\n Z_{\s K} \bigg) \dd x^\r \otimes \dd x^{\m \n K}~,
\eeq
and this is $O(\ap)$ since
\beq\notag
[\nabla_\m,\nabla_\lambda] Z_r \= (\Gamma_\lambda{}^\s{}_\m - \Gamma_\m{}^\s{}_\lambda) \nabla_\s Z_r \= H_\lambda{}^\s{}_\m \nabla_\s Z_r \=  O(\ap)~,
\eeq
on holomorphic indices, and $\partial_\nabla R = 0$ implies
\beq\notag
\nabla_\m R_\n{}^{\s \lambda}{}_\r  - \nabla_\n R_\m{}^{\s \lambda}{}_\r \= 0~.
\eeq
Here we recall that by the definition of $D$, the operator $\nabla$ acts as the Chern connection on $\s, \lambda, \r$ indices and as the exterior derivative on the form indices $\m, \n$. Altogether, \eqref{DD-direct} becomes
\beq\notag
D^2 \begin{bmatrix} Z \\ \aa \\ \D \end{bmatrix} \= \begin{bmatrix} 0 \\  0 \\ \tilde{\mathcal{H}} \partial Z + \partial( \tilde{\mathcal{H}} Z) + \ap (\tilde{\cF}^* \tilde{\cF} Z + \cR_2 \nabla \partial Z + \partial (\cR_2 \nabla Z) )\, \end{bmatrix} + \begin{bmatrix} O(\ap^2) \\ O(\ap) \\ O(\ap^2) \end{bmatrix}~.
\eeq
Next, we compute
\beq\notag
\tilde{\mathcal{H}} \partial Z + \partial( \tilde{\mathcal{H}} Z) \= g^{\r \tb} \partial_\n H_{\m \sb \tb} \, \partial_\r \otimes \dd x^{\n \m} \wedge Z^{\sb}~,
\eeq
and
\beq\notag
\tilde{\cF}^* \tilde{\cF} Z \= g^{\r \sb} \, {\rm Tr} \, F_{\n \sb} F_{\m \tb} \, \partial_\r \otimes dx^{\n \m} \wedge Z^{\tb}~.
\eeq
Using $\partial_\nabla R =0$, we compute
\beq\notag
[\cR_2 \nabla] \partial Z + \partial ([\cR_2 \nabla] Z) \=\! - g^{\r \sb} R_{\m \sb \tb }{}^{\lb} [\nabla_\n, \nabla_{\lb}] Z^{\tb}{}_{K} \, \partial_\r \otimes \dd x^{\n \m K}~,
\eeq
which becomes
\beq\notag
[\cR_2 \nabla] \partial Z + \partial ([\cR_2 \nabla] Z)\=\! - g^{\r \sb} {\rm Tr} \, ( R_{\n \sb} R_{\m \tb} )\, \partial_\r \otimes dx^{\n \m} \wedge Z^{\tb} + O(\ap)~,
\eeq
after using $R_{\n \tb \lb}{}^\rb = R_{\n \lb \tb}{}^\rb + O(\ap)$. These terms can be combined to give 
\beq\notag
D^2 \begin{bmatrix} Z \\ \aa \\ \D \end{bmatrix} \= \begin{bmatrix} 0 \\  0 \\ \half (-\partial H - \frac{\ap}{2} {\rm Tr} \, F \wedge F + \frac{\ap}{2} {\rm Tr} \, R \wedge R)_{\n \m}{}^\r{}_{\tb} \, \partial_\r \otimes dx^{\n \m} \wedge Z^{\tb} \end{bmatrix} + \begin{bmatrix} O(\ap^2) \\ O(\ap) \\ O(\ap^2) \end{bmatrix}.
\eeq
Using the heterotic Bianchi identity gives the result (\ref{DD-identity}).

\newpage

\section{The heterotic Hermitian-Yang-Mills correspondence}
We are now in a position to formulate the operator $\cD = D + \Dbar$ and compute its curvature. It being holomorphic and traceless with respect to the metric on $X$ will be equivalent to the heterotic  supersymmetry conditions and Bianchi identity to first order in $\ap$.

\subsection{The operator \texorpdfstring{$\cD = D + \Dbar$}{cD} and its curvature}
\label{s:Fieldstrength}
Let $q \in \Gamma(Q)$, which we write as before in terms of $Z \in \Omega^{1,0}$, $\aa \in \Gamma({\rm End} \, E)$ and $\D \in T^{1,0}$. We start by computing
\beq\notag
\begin{split}
 D \Dbar q \=&
\begin{bmatrix}
\partial_\nabla + \ap(\cR_1 \nabla) & 0 & 0 \\
\tilde{\cF} & \partial_\nabla & 0 \\
\tilde{\cH} + \ap (\cR_2 \nabla) & \ap \tilde{\cF}^* & \partial_\nabla + \ap (\cR_3 \nabla)
\end{bmatrix}
\begin{bmatrix} \delb Z + \ap \cF^* \aa + \cH \D + \ap (\cR \nabla) \D \\ \delb \aa + \cF \D \\ \delb \D
\end{bmatrix} \\
\=& \begin{bmatrix} {\rm I} \\ {\rm II} \\{\rm III} \end{bmatrix}~.
\end{split}
\eeq
Next, we compute
\bea
\Dbar D q &=&
\begin{bmatrix}
\delb & \ap \cF^* & \cH + \ap (\cR \nabla) \\
0 & \delb & \cF \\
0 & 0 & \delb
\end{bmatrix} 
\begin{bmatrix}
\partial_\nabla Z + \ap (\cR_1 \nabla) Z\\
\tilde{\cF} Z + \partial_\nabla \aa \\
\tilde{\cH} Z + \ap (\cR_2 \nabla) Z + \ap \tilde{\cF}^* \aa + \partial_\nabla \D + \ap (\cR_3 \nabla) \D
\end{bmatrix}
\nonumber\\
&=& \begin{bmatrix}
{\rm IV} \\
{\rm V} \\
{\rm VI}
\end{bmatrix}~.
\nonumber
\eea
The hermitian Yang-Mills condition is equivalent to evaluating $g^{\m\nb} [D_\m, \Dbar_\nb] q$, which corresponds to taking the difference of the above equations and tracing with respect to the metric on $X$. In appendix \sref{app:trace} we give the details, summarise the main points here.

\begin{itemize}
\item From $(I)+(IV)$ we find three conditions. The first is:
\beq\notag
 \hat{R}_{\r \sb}{}^\m{}_\m Z^{\sb} + g^{\m \nb} \big(i \partial \delb \omega + \frac{\ap}{2} {\rm Tr} \, R \wedge R - \frac{\ap}{2} {\rm Tr} \, F \wedge F  \big)_{\r \nb \m \sb} Z^{\sb} \= O(\ap^2) ~.
\eeq
This is $O(\ap^2)$ by the Bianchi identity and the Bismut Ricci-flat condition. The second condition vanishes via the Yang-Mills equation of motion \eqref{EOMF}, \eqref{YM-torsion}
 \beq
 \ap g^{\m \nb} \bigg[  {\rm Tr} \, \nabla_\m F_{\r \nb} \aa \bigg] \= O(\ap^2)~.
\eeq
Finally, the third condition is
\beq\label{Omega-V1}
g^{\m \nb} \nabla_\m H_{\s \r \nb} \D^\s - \ap g^{\m \nb} R_{\r \nb}{}^\kappa{}_\s [\nabla_\m, \nabla_\kappa] \D^\s - \ap g^{\m \nb} \nabla_\m R_{\r \nb}{}^\kappa{}_\s \nabla_\kappa \D^\s \= O(\ap^2)~,
\eeq
which holds due to the $H$-field equation of motion $\nabla^\nb H_\nb = 0$ \eqref{EOMH}, the condition for the holomorphic commutator $[\nabla_\mu,\nabla_\kappa] \Delta^\sigma=O(H)=O(\ap)$, and using the non--\K symmetries \eqref{nonK-curv-sym3}  and $g^{\m \nb} R_{\m \nb}{}^r{}_\ell = O(\ap)$ by (\ref{ddbar-omega2curv}). 

\item We next compute $ (II) + (V)$. We find two independent conditions. The first, 
\beq
g^{\m \nb} (-\nabla_{\nb} F_\m{}^\s Z_\s - F_{\t \nb} H_\m{}^{\t\s} Z_\s + \ap F_{\s \nb} R_\m{}^\s{}^{\r \lb} \nabla_{\lb} Z_\r) \= \cO(\ap)~,
\eeq
by the Yang-Mills equation of motion \eqref{EOMF}, followed by  the second
\beq \label{gauge-comps-a-terms}
g^{\m \nb} ( F_{\m \nb} \aa - \aa F_{\m \nb} + \ap  ({\rm Tr} \, F_{\r \nb} \aa ) F_\m{}^\r - \ap ({\rm Tr} \, F_\m{}^\r \aa) F_{\r \nb} ) \= 0~,
\eeq
 by the Hermitian-Yang-Mills equation $\o^2 F = 0$.

 \item Finally, we compute $(III)+(VI)$:
 \beq
g^{\m \nb} [D_\m, D_{\nb}] q^\r \=\! - g^{\m \nb} \nabla_{\nb} H_{\m \sb}{}^\r Z^{\sb} + \ap g^{\m \nb} R_\m{}^{\r}{}_{\sb}{}^{\kb} [\nabla_{\nb},\nabla_{\kb}] Z^{\sb} + \ap g^{\m \nb} \nabla_{\nb} R_\m{}^{\r}{}_{\sb}{}^{\kb} \nabla_{\kb} Z^{\sb} + O(\ap^2)~.  \nonumber
\eeq
This vanishes using the same data as what went into the vanishing of  \eqref{Omega-V1}.

We also have 
\beq
 \ap(g^{\m \nb} H_\m{}^{p \r} {\rm Tr} \, F_{p \nb} \aa - {\rm Tr} \, \nabla^\m F_\m{}^\r \aa) \= O(\ap^2)~.
\eeq
which vanishes  by the Yang-Mills equation with torsion \eqref{YM-torsion}.

Finally, the last condition uses the identities \eqref{ddbar-omega2curv}, \eqref{bianchi-with-torsion}, \eqref{nonK-curv-sym3} as well as  $g^{\m \nb} F_{\m \nb}=O(\ap)$ and $g^{\m \nb} R_{\m \nb}{}^\lambda{}_\s = O(\ap)$ to obtain
\bea\label{1formcomps-Zterms}
-\hat{R}_\s{}^{\r \m}{}_\m \D^\s +  g^{\m \nb} g^{\r \tb} \big( - \ii \partial \delb \omega  + \frac{\ap}{2} {\rm Tr} \, F \wedge F - \frac{\ap}{2} {\rm Tr} \, R \wedge R \big)_{\m \nb \tb \s} \D^\s \= O(\ap^2) \nonumber~,
\eea
which vanishes by the heterotic Bianchi identity and the Bismut connection being Ricci flat.

\end{itemize}

\subsection{The main result}

In \sref{s:Fieldstrength}  we found that if we assume supersymmetry and the Bianchi identity (which in turn is equivalent to the equations of motion to first order in $\ap$) then the operator $\cD$ is holomorphic and HYM. We now demonstrate the converse and so the desired equivalence. 
\begin{enumerate}
 \item  If $\mathcal{F}_{\mathcal{D}}^{0,2}=0$, then $\Dbar^2=0$ and \eqref{Dbar-squared-final} implies the heterotic Bianchi identity \eqref{bianchi-1storder}. 

\item Consider the matrix eqution $ \omega^2\mathcal{F}_{\mathcal{D}} =0$.  It has a  middle component which is precisely \eqref{gauge-comps-a-terms} and so
\beq 
[ g^{\m \nb} F_{\m \nb}, \aa ] \= O(\ap), \quad {\rm for} \ {\rm any} \ \aa \in \Gamma({\rm End} \, E)~.
\eeq
Therefore $g^{\m \nb} F_{\m \nb} = \lambda \, {\rm id} + O(\ap)$, and taking the trace and integrating over $X$ gives the first Chern class $c_1(E)$, which is zero by assumption. Hence $g^{\m \nb} F_{\m \nb} = O(\ap)$. 

\item  Lastly, consider again the matrix $\o^2 \mathcal{F}_{\mathcal{D}} =0$. Its top-left  component  is \eqref{1formcomps-Zterms}, which implies
\beq
\hat{R}_{\r \sb}{}^\m{}_\m \= O(\ap^2)~,
\eeq
where $\hat{R}$ is the curvature of the Bismut connection. Similarly, by \eqref{Omega-V1}, we obtain $\nabla^{\nb} H_{\nb \r \s}=O(\ap^2)$ which implies
\beq
\hat{R}_{\r\t}{}^\m{}_\m \= O(\ap^2)~.
\eeq

\end{enumerate}
We have linked the heterotic Hermitian-Yang-Mills equations \eqref{eq-heterotic-HYM0} with the heterotic Bianchi identity and the equations
\beq \label{bismut-ricci-flat}
g^{\m \nb} F_{\m \nb} \= O(\ap), \quad \hat{R}_{\r \tb}{}^\m{}_\m = O(\ap^2), \quad \hat{R}_{\r \beta}{}^\m{}_\m \= O(\ap^2)~.
\eeq
These are almost the equations from supersymmetry. Supersymmetric solutions to the equations of motion satisfy
\beq \label{torsion-constraint}
H_\m{}^\m{}_\r \= \partial_\r \log |\Omega|_\omega + O(\ap^2)~, \qquad H^\m{}_{\m \bar{\r}} \= \partial_{\bar{\r}} \log|\Omega|_\omega + O(\ap^2)~.
\eeq
Here $\Omega$ is a nowhere vanishing holomorphic $(3,0)$ form on the Calabi-Yau threefold $X$. The condition \eqref{torsion-constraint} is well-known to be equivalent to the condition from Strominger's paper \cite{Strominger:1986uh}; see e.g. \cite{ppz18}.

We now explain why \eqref{bismut-ricci-flat} does in fact imply \eqref{torsion-constraint}. A similar argument appears in \cite{Strominger:1986uh}. First, we write the identities \eqref{strominger-bismut1}, \eqref{strominger-bismut2} on the vanishing Bismut connection as
\bea \label{eq:hatzero1}
\partial_\r H_\m{}^\m{}_{\tb} - \partial_{\tb} H_\m{}^\m{}_\r + \partial_\r \partial_{\tb} \log |\Omega|_\omega^2 \= \hat{R}_{\r \tb}{}^\m{}_\m &=& O(\ap^2)~, \nonumber\\
\partial_\r H_\m{}^\m{}_\beta - \partial_\beta H_\m{}^\m{}_\r \= \hat{R}_{\r \beta}{}^\m{}_\m &=& O(\ap^2)~.
\eea
Here we used the identity for the Chern-Ricci curvature
\beq\notag
R_{\r \tb}{}^\m{}_\m \= \partial_{\tb} \partial_\r \log | \Omega |^2_\omega~,
\eeq
which uses that $\Omega$ is a holomorphic volume form. We now let $\tau \in \Omega^{1,0}(X)$ be given by
\beq\notag
\tau \= H_\m{}^\m{}_\r dx^\r~,
\eeq
and rewrite \eqref{eq:hatzero1} in terms of $\tau$ to obtain
\beq \label{bismut-ricci-flat-11}
\begin{split}
 -2 \ii \partial \delb \log |\Omega|_\omega &\= \ii \delb \tau - i \partial \bar{\tau} + O(\ap^2)~, \\
\partial \tau &\= O(\ap^2)~.
\end{split}
\eeq
Our assumption is that $g$ is Ricci--flat and \K at zeroth order so that
\beq\notag
\tau \= \ap \tau_1 + \ap^2 \tau_2 + \dots~.
\eeq
Since $\delb \bar{\tau} = O(\ap^2)$, it follows that $\bar{\tau}_1 \in \Omega^{0,1}(X)$ is $\delb$-closed. The background $X$ is a simply-connected threefold, so $h^{0,1}_{\delb}(X)=0$. Therefore $\bar{\tau}_1 = - \delb \bar{f}$, and we can write
\beq\notag
\tau_1 \=\! - \partial f~, \qquad \delb \tau_1 \= \partial \delb f~.
\eeq
Therefore
\beq\notag
\ii \delb \tau  + \ii \partial \bar{\tau} \= \ii \partial \delb (f - \bar{f}) + O(\ap^2)~,
\eeq
and using index notation this is
\beq \label{div-H-typestuff2}
- \nabla_\r H_\m{}^\m{}_{\tb} - \nabla_{\tb} H_\m{}^\m{}_\r \= \partial_\r \partial_{\tb} (f-\bar{f}) + O(\ap^2)~.\\[5pt]
\eeq
It turns out that
\beq \label{d-dagger-omega-twice}
g^{\r \tb}( \nabla_\r H_\m{}^\m{}_{\tb} + \nabla_{\tb} H_\m{}^\m{}_\r) \=0~.\\[5pt]
\eeq
One can derive \eqref{d-dagger-omega-twice} by taking various traces of the non-\K symmetries \eqref{nonK-curv-sym3}. We see that $g^{\r \tb} \partial_\r \partial_{\tb}(f-\bar{f}) = O(\ap^2)$. Note that $g^{\r \tb}\partial_\r \partial_{\tb}$ is not the Laplacian of $g$ (since $g$ is non-\K), however we can use general elliptic PDE theory to estimate $f$. For this, we first  shift $f$ by a constant such that $\int_X f =0$. The Schauder estimates for elliptic operators imply
\beq\notag
\| f - \bar{f} \|_{C^{2,\gamma}} ~\leq~ C \| g^{\r \tb}\partial_\r \partial_{\tb} (f-\bar{f}) \|_{C^{\gamma}} \leq C |\ap^2|~.\\[5pt]
\eeq
It follows that the right-hand side of \eqref{div-H-typestuff2} is of order $\ap^2$. 

\par We continue the argument which identifies $f$ with $-\log |\Omega|_\omega$. For this, we combine
\beq\notag
 \ii \delb \tau - \ii \partial \bar{\tau} \= \ii \partial \delb (f+\bar{f}) + O(\ap^2)~,
\eeq
with the $(1,1)$-part of the Bismut-Ricci flat identity \eqref{bismut-ricci-flat-11} to obtain
\beq
 \ii \partial \delb (f + \bar{f} + 2 \log |\Omega|_\omega) \= O(\ap^2)~.
\eeq
As before, PDE estimates imply that
\beq\notag
f + \bar{f} + 2 \log |\Omega|_\omega \= {\rm const} + O(\ap^2)~.
\eeq
Thus
\beq\notag
\partial f \=\!  -\partial \log |\Omega|_\omega + O(\ap^2)~.
\eeq
and $\tau = \partial \log |\Omega|_\omega + O(\ap^2)$. This proves the supersymmetry constraints \eqref{torsion-constraint}.

By a well-known argument (e.g. \cite{amp24, LopesCardoso:2003dvb, Martelli:2010jx}), solutions to the supersymmetry equations \eqref{torsion-constraint} solve the equations of motion where the action is given at first order in $\ap$, after setting the dilaton to be $\Phi = -\half \log |\Omega|_\omega$ .
 
 Thus we end up with the promised equivalence between the heterotic supersymmetry conditions \eqref{susy-1storder} and Bianchi identity \eqref{bianchi-1storder} and the operator $\cD$ on the bundle $Q$ being holomorphic and Hermitian-Yang-Mills \eqref{eq-heterotic-HYM0}.

\section{ Conclusion}
To summarize, we have demonstrated that, to suitable order in $\alpha'$, the equations
\bea \label{eq:regsusy}
& \ F^{0,2}\=0~, \qquad \ii \partial \bar{\partial} \omega \= \frac{\ap}{2} ({\rm Tr} \, F \wedge F - {\rm Tr} \, R \wedge R) \nonumber~,\\
& \omega^2 F \= 0~, \quad d (|\Omega|_\omega \omega^2)\=0~,
\eea
can be expressed on the bundle $(Q,\cD) \rightarrow X$ with 
\beq \notag
Q=(T^{1,0}X)^* \oplus {\rm End} \, E \oplus T^{1,0} X
\eeq
as the heterotic Hermitian-Yang-Mills equations
\beq \label{eq:heterotic-hym3}
\mathcal{F}_{\mathcal{D}}^{0,2}\=0~, \quad \mathcal{F}_{\mathcal{D}}^{2,0}\=0~, \quad \o^2 \mathcal{F}_{\mathcal{D}}  \= 0~.
\eeq
This can be compared to the calculations in \cite{delaossa:2017restrictions,garcia2023}, where $Q$ includes an ${\rm End} \, T^{1,0}X$ summand related to a spurious connection $\theta$ on $T^{1,0}X$. The heterotic Hermitian-Yang-Mills equation is peculiar as $\mathcal{D}$ is not a connection due to extra terms of order $\ap$. It is a differential operator of order 1 acting on $\Omega^{p,q}(Q)$, and $\bar{D}: \Omega^{0,q}(Q) \rightarrow \Omega^{0,q+1}(Q)$ satisfies $\bar{D}^2=0$. In upcoming work we will show that this operator is associated to an elliptic complex, in addition to many other nice mathematical properties. Note that since $\mathcal{D}$ is not a covariant derivative, the operator $\mathcal{F}_{\mathcal{D}}= \mathcal{D}^2$ is not a curvature in the traditional sense. 

For the standard Hermitian-Yang-Mills equation $\omega^2 F = 0$ for the curvature of the Chern connection on a holomorphic vector bundle $E \rightarrow X$, the solvability is determined by the Donaldson-Uhlenbeck-Yau theorem \cite{0529.53018,uhlenbeckyau} (see \cite{0664.53011} for the non-\K version). The theorem states that the existence of a Hermitian-Yang-Mills metric is equivalent to the polystability of $E \rightarrow (X,\omega)$. It does not apply here as $(Q,\cD)$ includes $\ap$ corrections to the holomorphic bundle structure and the metric $\omega$ is an unknown which appears in both the wedge $\omega^2$ and the operator $\bar{D}$, but we wonder whether an analog to the Donaldson-Uhlenbeck-Yau theorem exists. There is a proposal \cite{garcia2023, garcia202b} in this direction with spurious modes included.

One approach for solving the standard Hermitian-Yang-Mills equations is by Donaldson heat flow. It would be interesting to see if a heterotic version of the Donaldson heat flow can be defined in this setting, and whether it is related to the anomaly flow \cite{amp24, ppz18,ppz18b}. This sort of reformulation of a non-K\"ahler geometric flow into a Hermitian-Yang-Mills type flow occurs for the pluriclosed flow \cite{garciastreetsjordan}.

Lastly, we note that the equivalence \eqref{eq:regsusy} between \eqref{eq:heterotic-hym3} holds up to $O(\ap^2)$ in the action. The action \eqref{eq:10daction} of heterotic supergravity is correct up to $O(\ap^3)$, and so there may be a heterotic Hermitian-Yang-Mills correspondence to next order in $\ap$. The supersymmetry equations \eqref{eq:regsusy} will receive extra corrections at this next order, and so the relevant non-\K complex geometry will be modified beyond conformally balanced metrics.

\bigskip
\subsection*{Acknowledgements}
SP is supported by a NSERC Discovery Grant. JM and ES would like to thank the mathematical research institute MATRIX in Australia where part of this research was performed for a lively and rewarding research environment. We thank Anthony Ashmore, Ruben Minasian, Savdeep Sethi and Charlie Stickland-Constable for many interesting discussions during this time. We also thank Xenia de la Ossa, Hannah De L\'azari, Mario Garcia-Fernandez, Ra\'ul Gonz\'alez Molina, Magdalena Larfors, Jason Lotay, Matthew Magill, Enrico Marchetto, Javier Murgas Ibarra, Roberto Rubio, Henrique S\'a Earp and Roberto Tellez for useful discussions.

\newpage
\appendix
\section{Heterotic supergravity}
\label{app:sugra}
We summarise the $\ap^2$--corrected heterotic supergravity action following the notation in \cite{Anguelova:2010ed}. 
The heterotic action is fixed by supersymmetry up to and including $\ap^2$ corrections. There is a nice field basis constructed  in \cite{Bergshoeff:1988nn} to order $\ap$  by supersymmetrising the Lorentz--Chern--Simons terms in $H$ and extended to $\ap^2$ in \cite{Bergshoeff:1989de}.   The action, is given by 
\begin{equation} 
S \= \frac{1}{2\kappa_{10}^2} \int\! \dd^{10\,}\! X \sqrt{g_{10}}\, e^{-2\Phi} \Big\{ \cR -
\half |H|^2  + 4 |\del \Phi|^2 - \frac{\ap}{4}\big( \tr |F|^2 {-} \tr |R(\Theta^+)|^2 \big) \Big\} + \cO(\ap^3)~.
\label{eq:10daction}
\end{equation}
Here $\kappa_{10}$ is the 10D Newton constant, $\cR$ is the Ricci scalar curvature, $H$ is the 3-form field strength, $\Phi$ is the 10D dilaton, $F = \dd A+ A^2$ with the trace in the action taken in the adjoint of the gauge group, and \hbox{$g_{10}=-\det(g_{MN})$}. The action and Bianchi identity for $H$ is unique up to field redefinitions; there is no ambiguity in the theory.

The quadratic curvature terms are
$$
\tr |F|^2 \= \half \tr F_{MN} F^{MN}~~~\text{and}~~~ \tr |R(\Th^+)|^2 ~=~
\half  R_{MN}{}^A{}_B(\Theta^+) R^{MN}{}^B{}_A(\Theta^+)~.
$$
Here $R$ is the curvature of the twisted connection
$$
\Theta^\pm_M{}^A{}_B \= \Theta_M{}^A{}_B \pm \half H_M{}^A{}_B~,
$$
where $\Theta_M$ is the Levi-Civita connection. The 3-form $H$ and the curvature forms are related through the 4-form Bianchi--identity
\beq\notag
\dd H \=\!- \frac{\ap}{4} \left( \tr F^2 - \tr R^2\, \right)~.
\eeq
The 3-form $H$ is also related to the hermitian form via
\beq \notag
H \= \ii (\del - \delb) \o~.
\eeq
Therefore the Bianchi identity in terms of the hermitian form is
\beq\label{eq:Bianchi2}
2\ii \del \delb \o \=  \frac{\ap}{4} \left( \tr F^2 - \tr R^2\, \right)~.
\eeq 
 
The equations of motion, correct up to and including second order in $\ap$ in the action, are given by
\bea
&& R_{MN}+ 2 \nabla_M \nabla_N\Phi - \frac{1}{4} H_{MAB} H_N{}^{AB}\nonumber \\
&&\qquad\qquad-
\frac{\ap}{4} \Big( \tr F_{MP} F_N{}^P - R_{MPAB}(\Th^+)R_N{}^{PAB}(\Th^+)\Big)  + \cO(\ap^3) \= 0~,\\[6pt]\label{EOMR}
&&\nabla^M(\ee^{-2\Phi} H_{MNP}) + \cO(\ap^3) \= 0~,\label{EOMH}\\[8pt]
&&\ccD^{\Bi\,M} (\ee^{-2\Phi} F_{MN}) + \cO(\ap^2) \= 0~.\label{EOMF}
\eea
\vskip5pt
Here $\ccD^\Bi = \nabla^\Bi + [A,\cdot]$,  with $\nabla^\Bi$ computed with respect to the Bismut connection $\Th^\Bi = \Th - \half H$ and $R_{MN}$ is the Ricci tensor. Note that the equation of motion for the gauge sector is only accurate modulo $\cO(\ap^2)$ corrections. This is due to the fact that the gauge fields couple to supergravity at $\cO(\ap)$ in the action \eqref{eq:10daction}.

\section{Identities from complex geometry}
This section mostly serves to establish notation and conventions, and we also derive some identities in complex geometry that are used in the main text. The equations derived here hold generally on a complex manifold and we do not use any identities from supersymmetry.

Let $X$ be a complex manifold with hermitian metric $\omega = i g_{\r \tb} dx^\r \wedge d \bar{x}^\beta$.  Our conventions for the components of differential forms in holomorphic coordinates is
\beq
\eta \= \frac{1}{p!q!} \, \eta_{i_1 \cdots i_p \overline{j_1} \cdots \overline{j_q}} \, dx^{i_1 \cdots i_p} \wedge d \bar{x}^{\overline{j_1} \cdots \overline{j_q}}, \quad \eta \in \Omega^{p,q}(X)~.
\eeq
The 3-form field strength associated to $\omega$ will be denoted $H = i (\partial - \delb) \omega$, so that in components
\beq
H \= \half H_{\bar{\r} \m \n} d \bar{z}^\r \wedge d z^\m \wedge d z^\n + \half H_{\r \mb \nb} d z^\r \wedge d \bar{z}^\m \wedge d \bar{z}^\n~,
\eeq
then
\beq
H_{\bar{\r} \m \n} \=\! - (\partial_\m g_{\n \bar{\r}} - \partial_\n g_{\m \bar{\r}})~.
\eeq
Since we do not assume that $g$ is a \K metric, the Levi-Civita connection no longer preserves the holomorphic structure, and so by default we will use $\nabla$ and $R$ to denote the Chern connection of $g$ on $T^{1,0}X$. Our conventions are
\bea
\nabla_\m V^\lambda &\=& \partial_\m V^\lambda + \Gamma_\m{}^\lambda{}_\s V^\s, \quad \Gamma_\m{}^\n{}_\rho \= \partial_\m g_{\rho \sb} g^{\sb \n} \nonumber\\
\nabla_{\mb} V^\lambda &\=& \partial_{\mb} V^\lambda~,
\eea
for any section $V \in \Gamma(T^{1,0}X)$ and
\beq \label{R-chern-conv}
R_{\m \bar{\lambda}}{}^\n{}_\rho \=\! - \partial_{\bar{\lambda}} \Gamma_{\m}{}^\n{}_\rho, \quad [\nabla_\m,\nabla_{\bar{\lambda}}] V^\r \= R_{\m \bar{\lambda}}{}^\r{}_\beta V^\beta~.
\eeq
Our conventions imply
\beq \label{R-chern-conv2}
[\nabla_\lambda, \nabla_{\mb}] V^{\bar{\r}} \=\! - R_{\lambda \mb \tb}{}^{\bar{\r}} V^{\tb}, \quad [\nabla_\lambda,\nabla_{\mb}] V_\r \=\!- R_{\lambda \mb}{}^\beta{}_\r V_\beta~.
\eeq
We also have the non-\K symmetries
\beq \label{nonK-curv-sym3}
R_{\m \nb \tb \r} \= R_{\m \tb \nb \r} - \nabla_\m H_{\r \tb \nb}~, \quad R_{\m \nb \tb \r} \= R_{\r \nb \tb \m} - \nabla_{\nb} H_{\tb \r \m}~,
\eeq
\beq \label{bianchi-with-torsion}
\nabla_\m R_{s \bar{\gamma} \sb \lambda} \= \nabla_s R_{\m \bar{\gamma} \sb \lambda} + H^r{}_{s \m} R_{r \bar{\gamma} \sb \lambda}, \quad \nabla_{\kb} R_{\m \nb \tb \r}  \= \nabla_{\nb} R_{\m \kb \tb  \r } + H^{\bar{r}}{}_{\nb \kb} R_{\m \bar{r} \tb \r}~.
\eeq
\beq
\overline{R_{\m \nb}{}^\r{}_\beta} \= R_{\n \mb \tb}{}^{\r}~.
\eeq
We will also use the Bismut connection $\hat{\nabla}$ (also known as the Strominger-Bismut connection \cite{bismut1989, Strominger:1986uh}). Compared to the Chern connection $\nabla$, the Bismut connection is given in holomorphic coordinates by
\beq \label{nabla-hat}
\hat{\nabla}_\kappa V^\r \= \nabla_\kappa V^\r + H^\r{}_{\kappa \lambda} V^\lambda~, \qquad \hat{\nabla}_{\kb} V^\r \= \partial_{\kb} V^\r - H_{\lambda \kb}{}^\r V^\lambda~.
\eeq
This connection was noticed by Yano \cite{YanoBook} to be the unique connection on $T^{1,0}X$ such that $\hat{\nabla} g = 0$ and
\beq
\hat{T}^\r{}_{\kappa \beta} \= \hat{\Gamma}_\kappa{}^\r{}_\beta - \hat{\Gamma}_\beta{}^\r{}_\kappa =\ H^\r{}_{\kappa \beta}, \quad \hat{T}^\r{}_{\kb \beta} \= \hat{\Gamma}_{\kb}{}^\r{}_\beta - \hat{\Gamma}_\beta{}^\r{}_{\kb} \= H^\r{}_{\kb \beta}~.
\eeq
The Bismut curvature satisfies
\bea 
\hat{R}_{\r \tb}{}^\m{}_\m &\=& \partial_\r H_\m{}^\m{}_{\tb} - \partial_{\tb} H_\m{}^\m{}_{\r} + R_{\r \tb}{}^\m{}_\m \label{strominger-bismut1}~, \\
 \hat{R}_{\r \beta}{}^\m{}_\m &\=& \partial_\r H_\m{}^\m{}_\beta - \partial_\beta H_\m{}^\m{}_\r~, \label{strominger-bismut2}
\eea
where $R_{\r \tb}{}^\m{}_\m$ is the Chern-Ricci curvature.

\par We will encounter various traces of the curvature of the Chern connection in our calculations, and we note here the following identity:
\beq \label{ddbar-omega2curv}
 g^{\m \nb} R_{\m \nb \tb \r}\=\! - g^{\m \nb} (i \partial \delb \omega)_{\m \nb \tb \r} - \hat{R}_{\r \tb}{}^\m{}_\m + g^{\s \bar{\rho}} g^{\m \nb}  H_{\s \nb \tb} H_{\bar{\rho} \m \r } ~.
\eeq
Our conventions are such that $g^{\m \nb} R_{\m \nb \tb \r}=0$ is the condition for the Chern connection to be a Hermitian-Yang-Mills connection on $T^{1,0}X$. We now give the derivation of (\ref{ddbar-omega2curv}). Direct calculation gives
\beq
(i \partial \delb \omega)_{\m \nb \r \tb} \=\! -\partial_\m \partial_{\nb} g_{\r \tb} + \partial_\r \partial_{\nb} g_{\m \tb} + \partial_\m \partial_{\tb} g_{\r \nb} - \partial_\r \partial_{\tb} g_{\m \nb}~,
\eeq
and substituting the Chern curvature tensor leads to
\beq
(\ii \partial \delb \omega)_{\m \nb \r \tb} \=  R_{\m \nb \tb \r} - R_{\r \nb \tb \m} - R_{\m \tb \nb \r} + R_{\r \tb \nb \m}  - g^{\s \bar{\rho}} H_{\bar{\rho} \m \r} H_{\s \nb \tb}~.
\eeq
Tracing by $g^{\m \nb}$ then gives
\beq
- g^{\m \nb} (i \partial \delb \omega)_{\m \nb \tb \r} \=  g^{\m \nb} R_{\m \nb \tb \r} - R_{\r}{}^\m{}_{\tb \m} - R_{\m \tb}{}^\m{}_{ \r} + R_{\r \tb}{}^\m{}_\m  - g^{\m \nb} g^{\s \bar{\rho}} H_{\bar{\rho} \m \r} H_{\s \nb \tb}~.
\eeq
Taking traces of \eqref{nonK-curv-sym3} gives
\beq
R_\r{}^\m{}_{\tb \m} \= R_{\r \tb}{}^\m{}_\m - \nabla_\r H_{\m \tb}{}^\m, \quad R_{\m \tb}{}^\m{}_\r \= R_{\r \tb}{}^\m{}_\m - \nabla_{\tb} H^\m{}_{\r \m}~.
\eeq
Therefore
\beq
- g^{\m \nb} (\ii \partial \delb \omega)_{\m \nb \tb \r} \=  g^{\m \nb} R_{\m \nb \tb \r} + \nabla_\r H^\m{}_{\m \tb} + \nabla_{\tb} H_\m{}^\m{}_{\r} - R_{\r \tb}{}^\m{}_\m - g^{\m \nb} g^{\s \bar{\rho}} H_{\bar{\rho} \m \r} H_{\s \nb \tb}~.
\eeq
We obtain (\ref{ddbar-omega2curv}) by substituting the trace of the Bismut curvature. 

\par Next, we note the following 3-form field divergence identity:
\beq \label{div-3form}
\nabla^{\kb} H_{\r \beta \kb} \=\!  -\hat{R}_{\r \beta}{}^\m{}_\m~,
\eeq
where $\nabla$ is the Chern connection and $\hat{R}$ is the curvature tensor of the Bismut connection. To derive \eqref{div-3form}, we start by differentiating $H$ with the Chern connection, which gives
\beq
\nabla_\r H_{\m \n \kb} \= \partial_\r H_{\m \n \kb} - \Gamma_\r{}^\s{}_{\m} H_{\s \n \kb} - \Gamma_\r{}^\s{}_{\n} H_{\m \s \kb}~,
\eeq
with $\Gamma_\m{}^\kappa{}_{\n}-\Gamma_{\n}{}^\kappa{}_{\m} = H_{\m}{}^\kappa{}_{\n}$. Thus
\bea
\nabla_\r H_{\m \n \kb} - \nabla_\m H_{\r \n \kb} &\=& - \partial_\r (\partial_\m g_{\n \kb} - \partial_\n g_{\m \kb})  - \Gamma_\r{}^\s{}_{\m} H_{\s \n \kb} - \Gamma_\r{}^\s{}_{\n} H_{\m \s \kb} - (\m \leftrightarrow \r)~,\nonumber\\
&\=& \partial_\n(\partial_\r  g_{\m \kb}- \partial_\m g_{\r \kb}) + H_{\r \m}{}^\s H_{\s \n \kb} - \Gamma_\r{}^\s{}_{\n} H_{\m \s \kb} + \Gamma_\m{}^\s{}_{\n} H_{\r \s \kb} \nonumber~,\\
&=& \nabla_\n H_{\m \r \kb} + H_{\r \m}{}^\s H_{\s \n \kb} + H_{\r \n}{}^\s H_{\m \s \kb} - H_{\m \n}{}^\s H_{\r \s \kb}~.
\eea
We now contract with $g^{\r \kb}$ and simplify to obtain
\beq
\nabla^{\kb} H_{\m \n \kb} \=\! - \partial_\m H_\r{}^\r{}_\n + \partial_\n H_\r{}^\r{}_\m~.
\eeq
The expression for Bismut curvature gives \eqref{div-3form}.

\par We conclude this section by stating our conventions for vector bundles. Let $E \rightarrow X$ be a holomorphic vector bundle. Let $A$ be a connection on $E$ with curvature $F \in \Omega^2({\rm End} \, E)$. We will write $F_{ij}$ for the 2-form components of $F$ and omit the ${\rm End} \, E$ indices so that $F_{ij}$ is matrix-valued. If $A$ is the Chern connection of a metric on $E$, then in a holomorphic frame $A^{0,1}=0$ and $F \in \Omega^{1,1}({\rm End} \, E)$. We note the identity
\beq \label{YM-torsion}
\nabla_{\tb} F^{\tb \m} - \nabla^\m (g^{\s \bar{\rho}} F_{\s \bar{\rho}}) -  F^{\tb \r} H_{\r \tb}{}^\m\= 0~,
\eeq
where $\nabla$ is the Chern connection. To see this, we start with the Bianchi identity $\delb F = 0$, which in components reads
\beq
\partial_{\tb} F_{\m \nb} - \partial_{\nb} F_{\m \tb} \= 0~.
\eeq
Converting the Dolbeault derivative to the Chern connection introduces torsion.
\beq
\nabla_{\tb} F_{\m \nb} - \nabla_{\nb} F_{\m \tb} + ( \Gamma_{\tb}{}^{\bar{\rho}}{}_{\nb}- \Gamma_{\nb}{}^{\bar{\rho}}{}_{\tb} ) F_{\m \bar{\rho}} \= 0~.
\eeq
In terms of $H$, this is
\beq
\nabla_{\tb} F_{\m \nb} - \nabla_{\nb} F_{\m \tb} - H_{\tb \nb}{}^{\bar{\rho}} F_{\m \bar{\rho}}\= 0~.
\eeq
Contracting with $g^{\m \tb}$ implies \eqref{YM-torsion}.

\newpage
\section{The trace of the curvature of \texorpdfstring{$\cD$}{cD}}
\label{app:trace}
Let $q \in \Gamma(Q)$, which we write as before in terms of $Z \in \Omega^{1,0}$, $\aa \in \Gamma({\rm End} \, E)$ and $\D \in T^{1,0}$. 
Recall from \sref{s:Fieldstrength} that
\bea
[D, \Dbar] q 
&\=& \begin{bmatrix} {\rm I } \\ {\rm II } \\{\rm III } \end{bmatrix}
- 
\begin{bmatrix} {\rm IV } \\ {\rm V } \\{\rm VI } \end{bmatrix}~,
\eea
with
\beq\notag
\begin{split}
   {\rm I} \=&  \partial_\nabla \delb Z + \ap \partial_\nabla (\cF^* \aa) + \partial_\nabla (\cH \D) + \ap \partial_\nabla ((\cR \nabla) \D ) \\[2pt]
  &+ \ap (\cR_1 \nabla) (\delb Z + \ap \cF^* \aa + \cH \D + \ap (\cR \nabla) \D)~, \\[7pt]
 {\rm II} \=&  \tilde{\cF} (\delb Z + \ap \cF^* \aa + \cH \D + \ap (\cR \nabla) \D) \\[2pt]
&+ \partial_\nabla \delb \aa + \partial_\nabla ( \cF \D)~, \\[7pt]
 {\rm III} \=& \tilde{\cH} (\delb Z + \ap \cF^* \aa + \cH \D + \ap (\cR \nabla) \D) \\[2pt]
&+ \ap (\cR_2 \nabla) (\delb Z + \ap \cF^* \aa + \cH \D + \ap (\cR \nabla) \D) \\[2pt]
& + \ap \tilde{\cF}^* \delb \aa + \ap \tilde{\cF}^* \cF \D + \partial_\nabla \delb \D + \ap (\cR_3 \nabla) \delb \D~.\\[2pt]
\end{split}
\eeq
and
\beq
\begin{split}
{\rm IV} \=& \delb \partial_\nabla Z + \ap \delb (\cR_1 \nabla) Z + \ap \cF^* \tilde{\cF} Z + \ap \cF^* \partial_\nabla \aa \\[2pt]
& + \cH (\tilde{\cH} Z + \ap (\cR_2 \nabla) Z + \ap \tilde{\cF}^* \aa + \partial_\nabla \D + \ap (\cR_3 \nabla) \D) \\[2pt]
& + \ap (\cR \nabla) (\tilde{\cH} Z + \ap (\cR_2 \nabla) Z + \ap \tilde{\cF}^* \aa + \partial_\nabla \D + \ap (\cR_3 \nabla)\D)~, \\[7pt]
{\rm V} \=&  \delb (\tilde{\cF} Z) + \delb \partial_\nabla \aa \\[2pt]
&+ \cF (\tilde{\cH} Z + \ap (\cR_2 \nabla) Z+ \ap \tilde{\cF}^* \aa + \partial_\nabla \D + \ap (\cR_3 \nabla) \D)~, \\[7pt]
{\rm VI} \=& \delb ( \tilde{\cH} Z + \ap (\cR_2 \nabla) Z+ \ap \tilde{\cF}^* \aa) + \delb \partial_\nabla \D + \ap \delb (\cR_3 \nabla) \D~.
\end{split}
\eeq

We evaluate $g^{\m\nb}([D,\Dbar] q)_\r$ using the equations (I)-(VI) above with respect to the metric  $g_{\m\nb}$ on $X$ assuming $q$ is a section of   $\O^{1,0}(Q)$. As the equations need to hold for any combination of fields $Z, \aa, \D$ this is sufficient.

\subsection*{The first row}
This amounts to checking the trace of the terms $(I)-(IV)$ above.

\smallskip

\par $\bullet$ $Z$ terms: set $\aa=0$ and $V=0$. Then 
\beq
([D_\m,D_{\nb}] q)_\r \dd x^\r 
\= [\nabla_\m, \nabla_{\nb}] Z - \ap \cF^*_{\nb} \tilde{\cF}_\m Z + \ap (\cR_1 \nabla)_\m \nabla_{\nb} Z - \ap \nabla_{\nb} (\cR_1 \nabla)_\m Z + O(\ap^2)
\eeq
After contracting by $g^{\m \nb}$, we obtain
\bea \label{Z-terms1}
g^{\m \nb} ([D_\m,D_{\nb}] q )_\r &\=&\! -g^{\m \nb} R_{\m \nb \sb \r} Z^{\sb} - \ap g^{\m \nb} {\rm Tr} \, F_{\r \nb} F_{\m \sb} Z^{\sb} \\
&&- \ap g^{\m \nb} R_\m{}^{\s \gamma}{}_\r [\nabla_\gamma, \nabla_{\nb}] Z_\s + \ap g^{\m \nb} \nabla_{\nb} R_\m{}^{\s \gamma}{}_\r \nabla_\gamma Z_\s + O(\ap^2)~. \nonumber
\eea
The first term is related to $i \partial \delb \omega$ via (\ref{ddbar-omega2curv}). The last term (involving $\nabla R$) is of order $O(\ap^2)$ since $g^{\m \nb} R_{\m \nb \bar{\gamma} \r} = O(\ap)$ and we may use the differential Bianchi identity for the Chern connection \eqref{bianchi-with-torsion}. Thus
\beq
\begin{split}
 g^{\m \nb} ([D_\m,D_{\nb}] q )_\r &\=  \hat{R}_{\r \sb}{}^\m{}_\m Z^{\sb} + g^{\m \nb} \bigg( (i \partial \delb \omega)_{\m \nb \sb \r}  - \ap {\rm Tr} \, F_{\r \nb} F_{\m \sb} \bigg) Z^{\sb} \\
 &\qquad\qquad + \ap g^{\m \nb} R_{\m}{}^{\beta \gamma}{}_\r R_{\gamma \nb \sb \beta}  Z^{\sb}  + O(\ap^2)~.
\end{split}
\eeq
Using \eqref{nonK-curv-sym3} to switch indices on $R$:
\beq
\begin{split}
 g^{\m \nb} R_{\m}{}^{\beta \gamma}{}_\r R_{\gamma \nb \sb \beta}  Z^{\sb} 
\=& g^{\m \nb} {\rm Tr} \, R_{\r \nb} R_{\m \sb} Z^{\sb} + O(\ap)~.
\end{split}
\eeq
Since $g^{\m \nb} F_{\m \nb} = O(\ap)$ and $g^{\m \nb} R_{\m \nb} = O(\ap)$, we obtain 
\beq \label{1formcomps-Zterms_b}
g^{\m \nb} ([D_\m,D_{\nb}] q )_\r \= \hat{R}_{\r \sb}{}^\m{}_\m Z^{\sb} + g^{\m \nb} \big(i \partial \delb \omega + \frac{\ap}{2}  {\rm Tr} \, R \wedge R - \frac{\ap}{2} {\rm Tr} \, F \wedge F  \big)_{\r \nb \m \sb} Z^{\sb}+ O(\ap^2)~,
\eeq
and this quantity is $O(\ap^2)$ by the heterotic Bianchi identity and the Bismut Ricci-flat condition.

\smallskip

\par $\bullet$ $\aa$ terms: set $V=0$ and $Z=0$. Then the $\Omega^{1,0}$ components are:
\beq
 ([D_\m,D_{\nb}] q)_\r \dd x^\r \= \ap \nabla_\m (\mathcal{F}^*_{\nb} \aa) - \ap \mathcal{F}^*_{\nb} \nabla_\m \aa + O(\ap^2)~.
\eeq
After contracting with $g^{\m \nb}$, this is
\beq
g^{\m \nb} ([D_\m,D_{\nb}] q)_\r \= \ap g^{\m \nb} \bigg[  {\rm Tr} \, \nabla_\m F_{\r \nb} \aa \bigg] + O(\ap^2)~.
\eeq
By the Yang-Mills equation with torsion \eqref{YM-torsion}, this is of order $O(\ap^2)$.

\smallskip

\par $\bullet$ $\D$ terms: set $Z=0$ and $\aa=0$. Extracting the $\Omega^{1,0}$ components from our previous calculations, we get:
\beq
([D_\m,D_{\nb}] q)_\r \dd x^\r \= \nabla_\m (\mathcal{H}_{\nb} \D) + \ap \nabla_\m (  (\mathcal{R} \nabla)_{\nb} \D) - \mathcal{H}_{\nb} \nabla_\m \D  - \ap (\mathcal{R} \nabla)_{\nb} \nabla_\m \D + O(\ap^2)~.
\eeq
Here we can use the Chern connection in the definition of $(\cR \nabla)$ instead of $\nabla^+$, as they differ by $O(\ap)$. We contract with $g^{\m \nb}$ to obtain
\beq \label{Omega-V1_b}
\begin{split}
 \ & g^{\m \nb} ([D_\m,D_{\nb}] s)_\r \\
\=& g^{\m \nb} \nabla_\m H_{\s \r \nb} \D^\s - \ap g^{\m \nb} R_{\r \nb}{}^\kappa{}_\s [\nabla_\m, \nabla_\kappa] \D^\s - \ap g^{\m \nb} \nabla_\m R_{\r \nb}{}^\kappa{}_\s \nabla_\kappa \D^\s \\
&\qquad + O(\ap^2)~. 
\end{split}
\eeq
The first term is zero since $\nabla^{\nb} H_{\s \r \nb}=0$ (\ref{div-3form}). Next, we note
\beq
R_{\r \nb}{}^\kappa{}_\s \nabla_\m \nabla_\kappa \D^\s - R_{\r \nb}{}^\kappa{}_\s \nabla_\kappa \nabla_\m \D^\s \= O(\ap)~.
\eeq
since $[\nabla_\m,\nabla_\kappa]\D^\s = -H^\beta{}_{\m \kappa} \nabla_\beta \D^\s$. Finally, non-\K symmetries \eqref{bianchi-with-torsion} give
\beq 
g^{\m \nb} \nabla_\m R_{\r \nb}{}^r{}_\s \nabla_r \D^\s \= g^{\m \nb} \nabla_\r R_{\m \nb}{}^\kappa{}_\s \nabla_\kappa \D^\s + O(\ap)~,
\eeq
and we can then use $g^{\m \nb} R_{\m \nb}{}^r{}_\ell = O(\ap)$ by (\ref{ddbar-omega2curv}). Therefore all terms in (\ref{Omega-V1}) are of order $O(\ap^2)$.

\smallskip

\subsection*{The second row}
This amounts to the term given by $II+ V$.

$\bullet$ $Z$ terms: set $\aa = 0$ and $V=0$. The ${\rm End} \, E$ components are
\beq
[D_\m,D_{\nb}]q = \tilde{\cF}_\m \delb_{\nb} Z - \delb_{\nb} (\tilde{\cF}_\m Z) - \cF_{\nb} \tilde{H}_\m Z - \ap \cF_{\nb} (\cR_2 \nabla)_\m Z~.
\eeq
After contracting by $g^{\m \nb}$ this becomes
\beq
g^{\m \nb}[D_\m,D_{\nb}]q = g^{\m \nb} (-\nabla_{\nb} F_\m{}^\s Z_\s - F_{\t \nb} H_\m{}^{\t\s} Z_\s + \ap F_{\s \nb} R_\m{}^\s{}^{\r \lb} \nabla_{\lb} Z_\r)~.
\eeq
The Yang-Mills equation with torsion (\ref{YM-torsion}) implies that the first two terms sum to zero. This leaves the last term, which is $O(\ap)$.

\smallskip
\par $\bullet$ $\aa$ terms: set $V=0$ and $Z=0$. The ${\rm End} \, E$ components are
\beq
[D_\m, D_{\nb}]q = \nabla_\m \nabla_{\nb} \aa - \nabla_{\nb} \nabla_\m \aa + \ap \tilde{\cF}_\m \cF^*_{\nb} \aa - \ap \cF_{\nb} \tilde{\cF}^*_\m \aa~,
\eeq
which after contraction by $g^{\m \nb}$ become
\beq \label{gauge-comps-a-terms_b}
g^{\m \nb} ( F_{\m \nb} \aa - \aa F_{\m \nb} + \ap  ({\rm Tr} \, F_{\r \nb} \aa ) F_\m{}^\r - \ap ({\rm Tr} \, F_\m{}^\r \aa) F_{\r \nb} )~.
\eeq
The first two terms are zero by the Hermitian-Yang-Mills equation, and the last two terms add to zero.

\smallskip
\par $\bullet$ $\D$ terms: set $Z=0$ and $\aa=0$. We obtain
\beq
[D_\m,D_{\nb}]q = \tilde{\cF}_\m \mathcal{H}_{\nb} \D + \ap \tilde{\cF}_\m (\cR \nabla)_{\nb} \D + \nabla_\m (\cF_{\nb} \D) - \cF_{\nb} \nabla_\m \D - \ap \cF_{\nb} (\cR_3 \nabla)_\m \D~,
\eeq
which after contraction by $g^{\m \nb}$ becomes
\beq
g^{\m \nb} (F_\m{}^\r \D^\t H_{\t \r \nb} - \ap F_\m{}^\r R_{\r \nb }{}^\s{}_\l \nabla_\s \D^\l + \nabla_\m F_{\r \nb} \D^\r + \ap F_{\t \nb} R_\r{}^{\t \s}{}_\m \nabla_\s \D^\r)~.
\eeq
The first and third term sum to zero by the Yang-Mills equation with torsion (\ref{YM-torsion}). 
\smallskip
\par Altogether, we see that the gauge field components of $[D_\m,D_{\nb}]q$ are of order $O(\ap)$. This is down an order of $\ap$ compared to the other components, which are of order $O(\ap^2)$, as expected. Recall, the gauge field is implicitly normalised with an $\ap$ in its anomaly free coupling to gravity in $d=10$.

\subsection*{The third row}
We check the third row, or the terms  $III + VI$.

\par $\bullet$ $Z$ terms: set $\aa = 0$ and $V=0$. The vector field components are 
\beq
([D_\m,D_{\nb}] q)^\r \partial_\r = \tilde{\mathcal{H}}_\m \partial_{\nb} Z + \ap (\cR_2 \nabla)_\m \partial_{\nb} Z - \partial_{\nb} (\tilde{\mathcal{H}}_\m Z) - \ap \partial_{\nb} (\cR_2 \nabla)_\m Z + O(\ap^2)
\eeq
which implies
\beq
g^{\m \nb} [D_\m, D_{\nb}] q^\r = - g^{\m \nb} \nabla_{\nb} H_{\m \sb}{}^\r Z^{\sb} + \ap g^{\m \nb} R_\m{}^{\r}{}_{\sb}{}^{\kb} [\nabla_{\nb},\nabla_{\kb}] Z^{\sb} + \ap g^{\m \nb} \nabla_{\nb} R_\m{}^{\r}{}_{\sb}{}^{\kb} \nabla_{\kb} Z^{\sb} + O(\ap^2).  \nonumber
\eeq
This is the negative conjugate of \eqref{Omega-V1} with $\r$ index raised and $\overline{\D^\s} = Z^{\sb}$, and so these terms vanish to the correct order as before.

\par $\bullet$ $\aa$ terms: set $Z = 0$ and $V=0$. The vector field components are
\beq
([D_\m,D_{\nb}] q)^\r \partial_\r =  \ap \tilde{\mathcal{H}}_\m \cF^*_{\nb} \aa + \ap \tilde{\cF}_\m^* \nabla_{\nb} \aa - \ap \nabla_{\nb} (\tilde{\cF}^*_\m \aa)+ O(\ap^2),
\eeq
which simplify to
\beq
g^{\m \nb} [D_\m, D_{\nb}] q^\r =  \ap(g^{\m \nb} H_\m{}^{p \r} {\rm Tr} \, F_{p \nb} \aa - {\rm Tr} \, \nabla^\m F_\m{}^\r \aa) + O(\ap^2).
\eeq
The remaining term is zero by the Yang-Mills equation with torsion \eqref{YM-torsion}.

\par $\bullet$ $\D$ terms: set $Z = 0$ and $\aa=0$. The vector field components are
\bea
& \ & ([D_\m,D_{\nb}] q)^\r \partial_\r \nonumber\\
&=& [\nabla_\m,\nabla_{\nb}] \D + \ap \tilde{\cF}^*_\m \cF_{\nb} \D + \ap (\cR_3 \nabla)_\m \delb_{\nb} \D - \ap \nabla_{\nb} (\cR_3 \nabla)_\m \D+ O(\ap^2). \nonumber
\eea
Therefore
\bea
g^{\m \nb} [D_\m, D_{\nb}] q^\r &=& g^{\m \nb} R_{\m \nb}{}^\r{}_\s \D^\s + \ap g^{\m \nb} {\rm Tr} \, F_\m{}^\r F_{\s \nb} \D^\s \nonumber\\
&&- \ap g^{\m \nb} R_{\m}{}^{\r \gamma}{}_\s [\nabla_\gamma, \nabla_{\nb}] \D^\s + \ap g^{\m \nb} \nabla_{\nb} R_\m{}^{\r \gamma}{}_\s \nabla_\gamma \D^\s + O(\ap^2)~.
\eea
Using \eqref{ddbar-omega2curv} and \eqref{bianchi-with-torsion} implies
\beq\notag
\begin{split}
 g^{\m \nb} [D_\m, D_{\nb}] q^\r \=&\! -\hat{R}_\s{}^{\r \m}{}_\m \D^\s - g^{\m \nb} g^{\r \tb} \bigg(  (\ii \partial \delb \omega)_{\m \nb \tb \s}  - \ap {\rm Tr} \, F_{\m \tb} F_{\s \nb} + \ap R_{\m \tb}{}^{\lambda}{}_\gamma R_{\lambda \nb}{}^\gamma{}_\s \bigg) \D^\s \\
 &\qquad + O(\ap^2)~.
\end{split}
\eeq
Using \eqref{nonK-curv-sym3}, $g^{\m \nb} F_{\m \nb}=O(\ap)$ and $g^{\m \nb} R_{\m \nb}{}^\lambda{}_\s = O(\ap)$, we obtain
\beq
\begin{split}
 g^{\m \nb} [D_\m, D_{\nb}] q^\r \=&\! -\hat{R}_\s{}^{\r \m}{}_\m \D^\s -  g^{\m \nb} g^{\r \tb} \big(  \ii \partial \delb \omega  - \frac{\ap}{2} {\rm Tr} \, F \wedge F + \frac{\ap}{2} {\rm Tr} \, R \wedge R \big)_{\m \nb \tb \s} \D^\s\\
 &\qquad + O(\ap^2)~, \nonumber
\end{split}
\eeq
which vanishes by the heterotic Bianchi identity and the Bismut connection being Ricci flat.

\newpage

\vskip50pt
\raggedright
\baselineskip=10pt
\bibliographystyle{JHEP.bst}

\providecommand{\href}[2]{#2}\begingroup\raggedright\endgroup

\end{document}